\documentclass[preprint]{aastex}

\input{psfig.sty}
\def\arcsec {$^{\prime \prime}$}

\def\etal   {{\it et~al.\/}}

\def\HII    {H~{\rm {II}}}
\def\kms    {~km~s$^{-1}$}
\def\mo     {{$M_{\odot}$}}

\begin{document}

\title{The DEEP Groth Strip Survey XII: 
The Metallicity of Field Galaxies at $0.26<z<0.82$
and the Evolution of the Luminosity-Metallicity Relation}

\author{Henry A. Kobulnicky}
\affil{Department of Physics \& Astronomy \\ 
University of Wyoming \\ Laramie, WY 82071 
\\ Electronic Mail: chipk@uwyo.edu}

\author{Christopher N.~A. Willmer\footnote{On 
leave from Observat\'orio Nacional, Rio de Janeiro, Brazil}, 
Andrew C. Phillips, 
David C. Koo, \\  S.~M. Faber, \& Benjamin J. Weiner}
\affil{University of California, Santa Cruz \\ Department of Astronomy \& Astrophysics \\ 
Santa Cruz, CA, 95064  \\ 
cnaw@ucolick.org, bjw@ucolick.org, koo@ucolick.org,  
phillips@ucolick.org, faber@ucolick.org}

\author{Vicki L. Sarajedini}
\affil{Department of Astronomy \\ University of Florida \\
Gainesville, FL 32611 \\  vicki@astro.ufl.edu}

\author{Luc Simard}
\affil{Canadian Astronomy Data Centre \\
Herzberg Institute of Astrophysics \\ National Research Council of Canada  \\ 
 luc.simard@nrc.ca }

\author{Nicole P. Vogt} 
\affil{Department of Astronomy \\ New Mexico State University \\
	P.O. Box 30001, Dept 4500 \\  Las Cruces, NM 88003-8001 \\ 
nicole@nmsu.edu}

\author{Final revised draft of 27 August 2003}



\vskip 1.cm

\begin{abstract}

Using spectroscopic data from the Deep Extragalactic Evolutionary
Probe (DEEP) Groth Strip survey (DGSS), we analyze the gas-phase
oxygen abundances in the warm ionized medium 
for 64 emission-line field galaxies in the redshift
range $0.26<z<0.82$.  These galaxies comprise a small subset selected
from among 693 objects in the DGSS.  They are chosen for chemical
analysis by virtue of having the strongest emission lines.  Oxygen
abundances relative to hydrogen range between $8.4<12+log(O/H)<9.0$
with typical internal plus systematic measurement uncertainties of
0.17 dex.  The 64 DGSS galaxies collectively exhibit an increase in
metallicity with B-band luminosity, i.e., an L-Z relation like that
seen among local galaxies.  Using the DGSS sample and local galaxy
samples for comparison, we searched for a ``second parameter'' which
might explain some of the dispersion seen in the L-Z relation.
Parameters such as galaxy color, emission line equivalent width, and
effective radius were explored but found to be uncorrelated with
residuals from the the mean L-Z relation.

Subsets of DGSS galaxies binned by redshift also exhibit L-Z
correlations with slopes and zero-points that evolve smoothly with
redshift.  DGSS galaxies in the highest redshift bin ($z=0.6-0.82$)
are brighter, on average, by $\sim1$ mag at fixed metallicity compared
to the lowest DGSS redshift bin ($z=0.26-0.40$) and brighter by up to
$\sim2.4$ mag compared to local ($z<0.1$) emission-line field
galaxies.  Alternatively, DGSS galaxies in the highest redshift bin
($z=0.6-0.82$) are, on average, 40\% (0.15 dex) more metal-poor at
fixed luminosity compared to local ($z<0.1$) emission-line field
galaxies. For $0.6<z<0.8$ galaxies, the offset from the local L-Z
relation is greatest for objects at the low-luminosity ($M_B>-19$) end
of the sample and vanishingly small for objects at the high-luminosity
end of the sample ($M_B\sim-22$).  We compare these data to simple
single-zone exponential-infall \textsc{P\'egase2} models which follow
the chemical and luminous evolution of galaxies from formation to
$z=0$.  A narrow range of model parameters can qualitatively produce
the slope of the L-Z relation and the observed evolution of slope and
zero-point with redshift when at least two of the following are true:
1) low-mass galaxies have lower effective chemical yields than massive
galaxies, 2) low-mass galaxies assemble on longer timescales than
massive galaxies, 3) low-mass galaxies began the assembly process at a
later epoch than massive galaxies.  The single-zone models do a
reasonable job of reproducing the observed evolution for the
low-luminosity galaxies ($M_B\sim -19$) in our sample, but fail to
predict the relative lack of evolution in the L-Z plane observed for
the most luminous galaxies ($M_B\sim -22$).  More realistic
multi-zone models will be required to explain the chemo-luminous
evolution of large galaxies.
  
\end{abstract}

\keywords{ISM: abundances --- ISM: \HII\ regions --- 
galaxies: abundances --- 
galaxies: fundamental parameters --- 
galaxies: evolution ---
galaxies: starburst }

\section{Metallicity as a Measure of Galaxy Evolution} 

Many recent research programs in galaxy evolution trace changes in
correlations between fundamental galaxy properties as a function of
cosmic epoch.  One such approach compares the number density and
luminosity function of galaxies at earlier times (Lilly
\etal\ 1995; Bershady \etal\ 1997; Sawicki, Lin, \& Yee 1997; Lin
\etal\ 1999) to the luminosity function in the nearby universe (Zucca
\etal\ 1997; Marzke \etal\ 1998; Norberg \etal\ 2002).  Evidence suggests 
an increase in the number density of small, blue galaxies at earlier
times with only a small amount of passive fading among the more
luminous, redder galaxies.  Other investigations compare the relation
between rotation velocity and luminosity (T-F relation; Tully \&
Fisher 1977) locally with that observed in more distant disk galaxies
(e.g., Forbes \etal\ 1995; Vogt \etal\ 1996, 1997; Simard \& Pritchet 1998).
Most results indicate that galaxies of a given rotational amplitude
appear 0.2-1.0 mag brighter at $z\sim1$, although Bershady \etal\
(1998) argue that the variation is significant for only the bluest
galaxies, and Kannappan, Fabricant \& Franx (2002) show that
Tully-Fisher residuals are strongly correlated with galaxy color.
Simard \etal\ (1999) find no evidence for evolution of the
size-magnitude relation among 190 field galaxies out to $z=1.1$ in the
Groth Strip (Groth 1994).  They show that selection effects favor high
surface brightness galaxies in existing surveys.  These biases may
mimic the appearance of luminosity evolution and/or surface brightness
evolution reported in many studies (Schade \etal\ 1996a, 1996b; Roche
\etal\ 1998; Lilly
\etal\ 1998).  Yet another line of investigation compares the
Fundamental Plane\footnote{The fundamental plane is the locus
populated by early type galaxies galaxies in the three-dimensional
parameter space of surface brightness, velocity dispersion, and size
(parameterized by effective radius) (Dressler \etal\ 1987; Djorgovski
\& Davis 1987).} correlations for early-type galaxies as a function of
epoch.  Recent results suggest a decrease in the mean mass-to-light
ratio with lookback time for elliptical galaxies (van Dokkum \& Franx
1996; Kelson \etal\ 1997).  In Paper X of this series, Im \etal\
(2001) found a brightening of 1-2 mag for early-type $L^*$ galaxies
out to $z\sim1$, and in Paper IX of this series, Gebhardt \etal\ find 
an increase in the surface brightness by a similar amount.

In this paper we explore the evolution of the correlation between
metallicity, $Z$, and total luminosity, i.e., the
luminosity-metallicity (L-Z) relation to which nearly all types of
{\it local} galaxies conform.  In early-type systems, absorption line
indices provide a measure of the mean stellar iron or magnesium
abundances (Faber 1973; Brodie \& Huchra 1991; Trager 1999) while in
spiral (Zaritsky, Kennicutt, \& Huchra 1994) and dwarf systems
(Lequeux \etal\ 1979; French 1980; Skillman, Kennicutt, \& Hodge 1989; Richer \&
McCall 1995) the \HII\ region oxygen abundance relative to hydrogen is
the basis of the metallicity measurement.  The luminosity-metallicity
relation (L-Z) can be used as a sensitive probe and consistency check
of galaxy evolution.  The metallicity of a galaxy can only increase
monotonically with time (unless large-scale infall of primordial gas
is invoked), while the luminosity may increase or decrease depending
on the instantaneous star formation rate.  Metallicity is less
sensitive to variations due to transient star formation events in a
galaxy's history.  Most models and observations of galaxies at earlier
epochs predict higher star formation rates and a larger fraction of
blue, star-forming galaxies in the past (e.g., Madau \etal\ 1996; Lilly
\etal\ 1996; Somerville \& Primack 1999).  Thus, a high or
intermediate-redshift galaxy sample ought to be systematically
displaced from the local sample in the luminosity-metallicity plane if
individual galaxies participate in these cosmic evolution processes.
However, if local effects such as the gravitational potential and
``feedback'' from winds and supernovae regulate the star formation and
chemical enrichment process, then the L-Z relation might be
independent of cosmic epoch.  Semi-analytic models of Somerville \&
Primack (1999), for example, show very little evolution of the L-Z
relation with redshift.  One goal of this paper is to provide data
capable of testing these alternative hypotheses.

At high redshift, absorption line surveys have been tracing the
chemical evolution of the Lyman-alpha forest and damped Lyman-alpha
systems for more than a decade (Bergeron \& Stasi\'nska 1986; Sargent
\etal\ 1988; Steidel 1990; Pettini \etal\ 1997; Lu, Sargent, \& Barlow
1997; Prochaska \& Wolf 1999), while several studies have recently begun to
trace the cosmic chemical evolution of {\it individual galaxies} with
known morphological and photometric properties.  Kobulnicky \& Zaritsky
(1999) measured \HII\ region oxygen and nitrogen abundances for a sample
of 14 compact star-forming galaxies with kinematically narrow emission
lines in the range $z=0.1-0.5$.  Their sample was found to conform to
the local L-Z relation, within observational uncertainties.  Carollo \&
Lilly (2001) studied 13 star-forming galaxies at $0.5<z<1$ from the
Canada-France Redshift Survey (CFRS; Lilly \etal\ 1995) and found no
significant evidence for evolution in the L-Z relationship out to $z=1$.
However, at larger redshifts of $z>2$, both Kobulnicky \& Koo (2000)
and Pettini \etal\ (2001) found that Lyman break galaxies with
metallicities between $0.1~Z_\odot<Z<0.8~Z_\odot$ are 2-4 magnitudes
more luminous than local galaxies of similar metallicity.  This
deviation from the local L-Z relation demonstrates that the
luminosity-to-metal ratio varies throughout a galaxy's lifetime and is a
potentially powerful diagnostic of its evolutionary state.

The DEEP (Deep Extragalactic Evolutionary Probe; Vogt \etal\ 2003;
Paper I) team has been assembling Keck spectroscopic data on galaxies
in the Groth Strip survey (Groth 1994) in order to study the evolution
of field galaxies.  Previous papers in this series include a study of
the evolution of the Fundamental Plane (FP) of early-type galaxies
(Paper IX; Gebhardt \etal\ 2003), the evolution of the luminosity
function of E/S0 galaxies (Paper X; Im \etal\ 2001), and the
structural parameters of Groth Strip galaxies (Simard \etal\ 2002;
Paper II).  In this paper, we explore the chemical properties of 64
star-forming emission-line field galaxies observed as part of the DEEP
Groth Strip Survey (DGSS).  We measure the interstellar medium oxygen
abundances using the the ratio of strong [O~II], [O~III], and H$\beta$
emission line equivalent widths (e.g., Pagel \etal\ 1979 as extended
by Kobulnicky \& Phillips 2003; hereafter KP03).  We analyze the degree of metal
enrichment as a function of redshift, luminosity, and other
fundamental parameters.  Throughout, we adopt a cosmology with
$H_0$=70 \kms\ Mpc$^{-1}$, $\Omega_m=0.3$, and $\Omega_\Lambda=0.7$.

\subsection{Local Comparison Galaxy Samples }

In order to establish whether DGSS galaxies systematically differ
from nearby galaxies in their chemical or luminous properties,
we require a suitable sample of local galaxies for comparison.  
For a comparable sample of local star-forming galaxies, we compiled
three sets of emission-line
galaxy spectra from the literature.  The first set, consisting of 16 objects,
comes from the 55-object spatially-integrated
spectroscopic galaxy atlas of Kennicutt
(1992b) supplemented by 6 objects from Kobulnicky, Kennicutt, \&
Pizagno (1999).  We refer to this as the K92 local sample.  For each
galaxy with detectable (S/N$>$8:1) [O~II]$\lambda$3727,
[O~III]$\lambda$4959, [O~III]$\lambda$5007, and H$\beta$ emission, we
measured the emission-line fluxes and equivalent widths in the same manner as
DGSS galaxies described below.    As
a second local sample, we selected the compilation of nearby field
galaxies from Jansen \etal\ (2000a,b; NFGS) with integrated
spectra, adopting their published
emission-line fluxes and equivalent widths.  The Nearby Field Galaxy
Survey is a collection of 200 local ($z<0.04$) morphologically diverse
galaxies selected from the CfA I redshift catalog (Huchra 1983) which
has the virtue of being nearly complete to the photographic limits of
the survey.  See Jansen \etal\ (2000a) for a discussion of the
differences between the NFGS and K92 samples.  Briefly, the K92
objects have a higher fraction of star-forming galaxies (objects with
strong emission lines) compared to the NFGS sample.  As a third local
sample, we chose the emission-line selected galaxies from the Kitt
Peak National Observatory Spectroscopic Survey (KISS; Salzer \etal\
2000).  KISS is a large-area objective prism survey of local
($z<0.09$) galaxies selected solely on the basis of strong $H\alpha$ emission lines.
Because of this selection criterion, 
the KISS galaxies are preferentially gas-rich star-forming systems.
KISS spectra were obtained using a fiber-fed spectrograph with
3\arcsec\ apertures, so, unlike the first two samples, the
spectra are not spatially integrated (i.e., do 
not include the entire galaxy).

\subsection{Target Selection and Observations}

The DEEP Groth Strip Survey (DGSS) consists of Keck spectroscopy over
the wavelength range $\sim$4400 \AA\ -- 9500 \AA\ obtained with the Low
Resolution Imaging Spectrometer (LRIS; Oke \etal\ 1995). 
The sample is magnitude limited at $I=23.5$ on the Vega system.
Spectra have
typical resolutions of $\sim$3--4 \AA.  Integration times ranged
from 3000 s to 18,000 s with a mean of around 6000 s.   The spatial
window of the spectral extraction was typically 1.5\arcsec.  A full
description appears in Vogt \etal\ (Paper I in this series; 2003).

We searched the spectroscopic database for galaxies in the Groth Strip
Survey with nebular emission lines suitable for chemical analysis.
Only galaxies where it was possible to measure all of the requisite
[O~II]$\lambda3727$, H$\beta$, [O~III]$\lambda$4959, and
[O~III]$\lambda$5007 lines were retained.   These criteria
necessarily exclude objects at redshifts of $z\lesssim0.26$ since the
requisite [O~II]$\lambda$3727 line falls below the blue limit of the
spectroscopic setup.  Likewise, objects with redshifts $z\gtrsim 0.82$ are
excluded because the [O~III]$\lambda$5007 line falls beyond the red
wavelength limit of the survey.  As of February 2002, there were 693
objects with Keck spectra with identified redshifts in the DEEP Groth
Strip Survey.  Of these 693 objects, 398 unique objects have
spectroscopic redshifts within the nominal limits
$0.26<z<0.82$.  The usable sample is further reduced because 23
candidates were positioned on the slitmask such that the [O~III] lines
fell off the red end of the spectral coverage.  An additional 49
objects were rejected because their position on the slitmask caused the
[O~II]$\lambda$3727 line to fall off the blue end of the spectral
coverage.  Furthermore, atmospheric $O_2$ absorption troughs between
6865 \AA\ -- 6920 \AA\ (the ``B band'') and between 7585 \AA\ --
$\sim7680$ \AA\ (the ``A band'') prohibit accurate measurement of
emission lines for objects in particular redshifts ranges.  We removed
14 objects from the sample in the redshift interval $0.410<z<0.426$
where the $H\beta$ line falls in the B band.  We removed 8 additional
objects in the redshift interval $0.56<z<0.58$ where $H\beta$ 
falls in the A band.  We removed 24 objects in the redshift interval
$0.52<z<0.54$ which places both [O~III] $\lambda$4959 and [O~III]
$\lambda$5007 in the A band.  The B band is sufficiently narrow that
either [O~III] $\lambda$4959 or [O~III] $\lambda$5007 is always
available for measurement.  Following this selection process, 276
objects remained.

An additional 210 objects were removed from the sample because the
$H\beta$ emission line was absent or too weak (S/N$<$8:1) for reliable
chemical determinations (Kobulnicky, Kennicutt, \& Pizagno 1999 for a
discussion of errors and uncertainties).   The spectra of objects
rejected due to a weak $H\beta$ line are usually dominated by stellar
continuum rather than nebular emission from star-forming regions.  Most
local early-type spirals and elliptical galaxies share these spectral
characteristics.  For these objects, H$\beta$ is seen in absorption
against the stellar spectrum of the galaxy.  Thus early type galaxies with
older stellar populations are preferentially rejected in favor of late
type galaxies with larger star formation rates.  There should, however,
be no metallicity bias introduced by rejecting these 210 objects since
we will compare the DGSS sample with a local sample selected in the
same manner: on the basis of $H\beta$, [O~III] $\lambda5007$ and 
[O~II] $\lambda3727$ emission lines measured with a signal-to-noise 
of 8:1 or better.

Objects with strong $H\beta$ but immeasurably weak [O~II] or [O~III]
lines present a data selection conundrum.  In principle, such objects
should be included in the sample to avoid introducing a metallicity
bias, but it is not possible to compute metallicities if the oxygen
lines are not detected with N/N of 8:1 or better.  Only 3 such objects were found in
the database.  For these objects, it appears that poor sky subtraction
caused the [O~III] features to be immeasurably weak.
Intrinsically-weak oxygen lines may be caused by either extremely high
($Z>2~Z_\odot$) or extremely low ($Z<0.05~Z_\odot$) metal content.  In
the latter case, oxygen lines are weak because of the lack of $O^{+}$
and $O^{++}$ ions.  However, in the local universe,
galaxies with extremely low intrinsic
abundances are under-luminous ($M_B>-15$), and no such faint galaxies
are included in our sample.  In the high-metallicity case, efficient
cooling decreases the mean collisional excitation level, reducing the
[O~III] line strengths.  However, the Balmer lines and [O~II] lines
are not strongly affected by reductions in electron
temperature.  It is unlikely that our rejection criteria bias the
sample by preferentially excluding either metal-poor or metal-rich
systems.

Of the original 693 objects, 66 galaxies remain.  These objects appear
in Table~1, along with their equatorial coordinates, redshift,
absolute blue magnitude, restframe $(B-V)_0$ color in the Johnson Vega
systems, half-light radius $R_{hl}$, and bulge fraction $B/T$ derived
from model fitting routines (Simard \etal\ 1999, 2002).  The restframe
($B-V)_0$ and ($B-R)_0$ colors used in this work were calculated
following the procedure of Lilly
\etal\ (1995), by interpolating the measured V-I color of DGSS galaxies
over a subset of Kinney \etal\ (1996) spectra, which were then used to
synthesize the different restframe colors.  Figure~1 shows their
redshifted spectra with major emission lines identified next to the
F814W HST greyscale images.  A cursory glance at the images reveals an
assortment of galaxy types, from small compact objects to spiral disks
and apparent mergers-in-progress.

In order to assess whether the 66 selected objects are representative
of the 398 galaxies with spectra in the $0.25<z<0.82$ redshift range,
Figure~\ref{hist} shows histograms of their morphological and
photometric properties.  The six panels show the redshift distribution
$z$, the absolute B magnitude $M_B$, the rest-frame color $(B-V)_0$,
the half-light radius $R_{hl}$, the bulge fraction $B/T$, and the
total asymmetry index, $R_T+R_A$\footnote{Definitions of DGSS
structural parameters may be found in Paper II of this series, Simard
\etal\ (2002).}.  Examination of Figure~\ref{hist} reveals that the 66
galaxies selected for chemical analysis are representative of the
entire DEEP sample in terms of their luminosities, redshift
distributions, sizes, and bulge fractions, but they are preferentially
bluer and more asymmetric than the sample as a whole.  This
disproportionate fraction of asymmetric galaxies with strong emission
lines may be understood as either 1) mergers which trigger star
formation and produce \HII\ regions, or 2) isolated galaxies dominated
by star-forming regions which give rise to an asymmetric morphology.
The possible systematic effects introduced by the selection criteria
will be discussed further below.

\section{Spectral Analysis}
\subsection{Emission Line Measurements and Uncertainties}

Spectra taken on different observing runs with different slitmasks
enable us to combine multiple spectra for most DGSS galaxies. After
wavelength calibrating each spectrum, we combined the available
2-dimensional spectra to produce a master spectrum with higher
signal-to-noise.  The spectra are not flux calibrated.  We manually
measured equivalent widths of the emission lines present in each
spectrum with the IRAF SPLOT routine using Gaussian fits.  The [O~II]
doublet, which is visibly resolved, was fit with two Gaussian
components and the sum recorded.  Table~1 lists the equivalent widths
and measurement uncertainties for each line.  Where [O~III]
$\lambda$4959 was below our nominal S/N threshold of 8:1, we
calculated the EW based on the strength of [O~III] $\lambda$5007
assuming an intrinsic ratio of 3:1.  The reported equivalent widths
are corrected to the rest frame using

\begin{equation}
EW_{rest} = EW_{observed}/(1+z)
\end{equation}

\noindent Associated uncertainties are computed taking
into account both the uncertainty on the line
strength and the continuum level placement using

\begin{equation}
\sigma_{EW} = \sqrt{ {{1}\over{C^2}}\sigma_L^2 +
{{L^2}\over{C^4}}\sigma_C^2 },
\end{equation}

\noindent where $L$, $C$, $\sigma_L$, and $\sigma_C$ are the line and
continuum levels in photons and their associated $1~\sigma$ uncertainties.  We
determine $\sigma_C$ manually by fitting the baseline regions
surrounding each emission line multiple times.  We adopt
$\sigma_L=\sqrt{12}\times RMS$ where 12 is the number of pixels summed
in a given emission line for this resolution, and RMS is the root-mean-squared
variations in an adjacent offline region of the spectrum.  Using this
empirical approach, the stated uncertainties implicitly include 
{\it internal} errors from one emission line relative to another
due to Poisson noise, sky background, sky subtraction, readnoise, and
flatfielding.  In nearly all cases, the continuum can be fit along a
substantial baseline region, so that $\sigma_C\ll\sigma_L$.
There are, however, additional uncertainties on the {\it absolute} values
of the equivalent widths due to uncertainties on the continuum placement,
particularly for the [O~II] $\lambda$3727 line, which are difficult to 
include in the error budget.  For several objects, the $H\beta$ emission
line falls on both the red and the blue halves of the spectrum so that
two independent measurements of $EW_{H\beta}$ are available.  In these five cases,
the two measurements differ by up to 20\%.  This difference
is understandable since the red and blue spectra were taken at different times
during a night, separated by a periods of slitmask realignment and
seeing variations. 
Small positional changes of the slitmask may alter the region of the galaxy
observed on the red versus blue spectra.  
Thus, the absolute values of the equivalent widths may be
uncertain by 20\%.  However, since we are primarily interested in the
oxygen abundances derived from the ratios of these equivalent widths,
the impact on the error budget is less severe when all emission
lines are measured on the same exposure.  This is rarely the case.  
[O~III] and $H\beta$ generally fall on the red spectrum and [O~II] 
on the blue spectrum. 
To account for differences in sky subtraction and slitmask placement 
on the red versus blue spectra, we therefore have assigned an uncertainty of 20\%
to the EW of [O~II] $\lambda$3727.  This error is not
reflected in column 13 of Table~1, but has been included in the oxygen abundance
computations that follow.

\subsection{AGN Contamination}

In the analysis that follows, accurate assessment of chemical
abundances in the warm ionized medium requires that the observed
emission lines arise in H~II regions powered by photoionization from
massive stars.  Non-thermal sources such as active galactic nuclei
(AGN) often produce emission-line spectra that superficially resemble
those of star-forming regions.  AGN must be identified as such because
blindly applying emission-line metallicity diagnostics calibrated from
H~II region photoionization models will produce erroneous
metallicities.

Traditionally, AGN can be distinguished from starbursts on the basis
of distinctive [N~II]/H$\alpha$, [S~II]/H$\alpha$, [O~I]/H$\alpha$,
[O~III]/H$\beta$, and [O~II]/[O~III] line ratios (see Heckman
1980; Baldwin, Phillips,
\& Terlevich 1981; Veilleux \& Osterbrock 1987).  However, some or
most of these diagnostic lines are unobservable in the DGSS and in the
current generation of ground-based optical surveys at increasingly
higher redshifts.  With these limitations in mind, Rola, Terlevich, \&
Terlevich (1997) have explored using ratios of $EW_{[O~II]}$ and
$EW_{H\beta}$ as substitute diagnostics for identifying AGN.  They 
distinguish AGN from normal star forming galaxies partly on the basis
of large $[Ne~III]
\lambda3868$ to [O~II] $\lambda3727$ ratios.  Two DGSS objects,
092-7832 and 203-3109 have
$EW_{[Ne~III]\lambda3826}/EW_{[O~II]\lambda3727}> 0.4$, compared to
$EW_{[Ne~III]\lambda3826}/EW_{[O~II]\lambda3727}\sim0.1-0.2$ for
normal star-forming objects (e.g., Kennicutt 1992).  Osterbrock (1989)
also notes the presence of enhanced $[Ne~III]\lambda3868$ as a common
signature in AGN and LINERs.  On this basis, we remove these two
$z\sim0.68$ objects from further analysis.

Following Rola, Terlevich, \& Terlevich (1997) we also explored the
diagnostic utility of enhanced $[O~II]\lambda3727$/$H\beta$ ratios as
indicators of AGN activity.  As shown by their Figure~5, there is
considerable overlap between bona-fide AGN and normal star-forming
galaxies.  Among the 66 DGSS galaxies, there are 8 objects with
$EW_{[O~II]}$/$EW_{H\beta}>3.5$ which might suggest the presence of
LINER or AGN activity.  None of these galaxies have measured
$[N~II]\lambda6584/H\alpha$ ratios, so more traditional diagnostics
are not available. However, we do not see enhanced
$[Ne~III]\lambda3826$/$[O~II]\lambda3727$ ratios in these galaxies, nor is there any
evidence for broad linewidths or large $[O~III]\lambda5007/H\beta$ ratios
that are also typical signatures of AGN.
Lacking any additional evidence to the contrary, we proceed under the
assumption that the 64 remaining galaxies in our sample are all
dominated by emission lines from star formation regions.
   
\subsection{Properties of Selected Galaxies} 

Figure~\ref{select} shows in greater detail the distribution of
magnitude, color, restframe $EW_{H\beta}$ and $H\beta$ luminosity for
the 64 remaining galaxies compared to the original set of 276 DGSS
objects with emission lines in the range $0.26<z<0.82$.  Symbols
distinguish objects by redshift: ``low'' (17 objects; $0.26<z<0.40$),
``intermediate'' (21 objects; $0.40<z<0.60$), and ``high'' (26
objects; $0.60<z<0.82$).  The ratio of comoving volumes in these three
redshift bins is approximately 1:3:4.6. Filled symbols denote galaxies
selected for chemical analysis, while open symbols denote the entire
set of galaxies in each redshift interval.  Points with error bars
denote the means and dispersions of each sample.  The lower row shows
$(B-V)_0$ color versus $M_B$.  Objects in the lowest redshift bin
exhibit a correlation between luminosity and $(B-V)_0$
color, while the samples in the latter two redshift bins do not. Note
also that the distribution of colors among the DGSS galaxies is
bimodal, especially in the higher two redshift bins.  A similar
bimodality is seen in the $0.2<z<1.1$ COMBO-17 survey (Bell \etal\
2003).  The DGSS galaxies selected for analysis are distributed evenly
within the bluer grouping of galaxies. The two highest redshift bins
contain a higher fraction of luminous blue objects which are absent in
the low-redshift bin.  In the high-redshift bin, the magnitude-limited
nature of the sample becomes obvious, as there are no galaxies fainter
than $M_B=-18$.  In all redshift bins, the selected galaxies are
preferentially those with the highest $EW_{H\beta}$ and the bluest
colors.  This effect is most significant in the highest redshift bin.
The correlation between blue magnitude and $H\beta$ luminosity in all
three redshift bins reflects the fact that the $H\beta$
luminosities were computed from $M_B$ and $EW_{H\beta}$.  
No extinction corrections have been applied to the data.
The distribution of equivalent widths indicates that the star
formation per unit luminosity is roughly similar for galaxies in the
two highest redshift intervals, but that the lowest redshift bin has
fewer objects with large EW.  This is most likely a volume effect
since there will be fewer extreme starburst objects in the smaller
volume between $z=0.26-0.40$ than  $z=0.40-0.60$ or  $z=0.60-0.80$.

\section{Analysis}
\subsection{Assessing the Physical and Chemical Properties of the Sample}

Oxygen is the most easily measured metal in \HII\ regions due to its
strong emission lines from multiple ionization species in the optical
portion of the spectrum.  Measurement of the oxygen abundance relative
to hydrogen is based upon the intensity ratio of the collisionally
excited [O~II]$\lambda$3727 and [O~III]$\lambda$4959,5007 lines
relative to Balmer series recombination lines (e.g., H$\beta$) using
standard analysis techniques (see Osterbrock 1989).  Even when the
physical conditions of the ionized gas such as electron temperature and
density cannot be measured, the ratio of strong forbidden oxygen
emission lines can still provide a measure of the overall metallicity
of the gas (the so-called strong line $R_{23}$ ratio method; Pagel
\etal\ 1979; Kobulnicky, Kennicutt, \& Pizagno 1999, hereafter KKP).  
However, in the
DGSS and other large spectroscopic surveys, relative emission-line
intensities are often not measured.  Only equivalent widths are
available.  In a companion paper to this one, Kobulnicky \& Phillips
(2003) demonstrate that the ratio of emission-line equivalent widths,
$EWR_{23}$, is a quantity comparable to $R_{23}$ and is suitable for
measuring oxygen abundances.  We adopt the $R_{23}$ calibration of
McGaugh (1991) relating the ratio of [O~III] $\lambda\lambda$ 4959,
5007, [O~II] $\lambda$3727, and $H\beta$ to the oxygen abundance relative
to hydrogen, O/H.   Although multiple prescriptions have been proposed
in the literature, (as reviewed in KKP and in
Figure~\ref{R23OH}), the exact choice is unimportant here since we are
only interested in {\it relative} abundances between local and distant
galaxy samples analyzed in the same manner.  Furthermore, the sample
includes only luminous galaxies ($M_B<-18$, with three exceptions) so
that, based on local analogs, they all are expected fall on the upper (metal-rich)
branch of the empirical strong-line calibrations.  The nine DGSS
galaxies with measurable [N~II] and H$\alpha$ lines all have
[N~II]/$H\alpha \gtrsim 0.1$, confirming that these objects belong on the
metal-rich branch of the calibration.\footnote{See, for instance,
Figure~7 of Kewley \& Dopita (2002) where only galaxies with
12+log(O/H)$>$8.4 have [N~II]/H$\alpha$ ratios $>$0.1.  Adopting the
less-plausible hypothesis that the DGSS galaxies lie on the metal-poor
branch of the double-valued strong-line $R_{23}$--$O/H$ relation would
require all the galaxies to have extremely low oxygen abundances,
$12+log(O/H)<8.0$.  Given that direct measurements of electron
temperatures and oxygen abundances in similarly luminous $z=0.4$
objects rule out the lower branch possibility (Kobulnicky \& Zaritsky
1999) we believe this upper-branch assumption to be generally valid.} 
To ensure that the NFGS and KISS comparison objects
are on the metal-rich branch, both local samples have been culled to
contain only objects with $F_{[N~II]\lambda6584}/F_{H\alpha}>0.15$.

In this paper, we compute oxygen abundances adopting the analytical
expressions of McGaugh (1991, 1998 as expressed in KKP) which are
based on fits to photoionization models for the metal-rich (upper)
branch of the $R_{23}$--O/H relation. In terms of the reddening
corrected line intensities, this relation is

\begin{eqnarray}
12+log(O/H) = 12 -2.939-0.2x-0.237x^2-0.305x^3-0.0283x^4-  \\
	y(0.0047-0.0221x-0.102x^2-0.0817x^3-0.00717x^4),
\end{eqnarray}

\noindent where

\begin{equation}
y \equiv\ \log(O_{32}) \equiv\ \log\Biggr(
	{{I_{[O~III]\lambda4959} + I_{[O~III] \lambda 5007}}
	\over{I_{[O~II]\lambda3727}}}\Biggr),
\end{equation}

\noindent and

\begin{equation}
x\equiv\ \log({R_{23}})\equiv\ \log\Biggr({{I_{[O~II] \lambda3727} + I_{[O~III]
\lambda4959} + I_{[O~III] \lambda 5007}}\over{I_{H\beta}}}\Biggr).
\end{equation}

\noindent Figure~\ref{R23OH} shows graphically the adopted relation between
$R_{23}$, $O_{32}$ and oxygen abundance.  Alternative relations from
the literature are shown for comparison.  A star marks the Orion
Nebula value (based on data of Walter, Dufour, \& Hester 1992) which
is in excellent agreement with the most recent solar oxygen abundance
measurement of $12+log(O/H)_\odot=8.7$ (Prieto, Lambert, \& Asplund
2001).\footnote{We have assumed the solar
oxygen abundance to be $12+log(O/H)_\odot\simeq$8.7 based on the new
solar oxygen abundance determination of Prieto, Lambert, \& Asplund
(2001).  This is 0.1-0.2 dex lower than the oft-adopted Anders \&
Grevesse (1989) value, but resolves the discrepancy between solar and
Orion Nebula oxygen abundances.  For the Orion Nebula,
$R_{23}=0.70\pm0.03$, $O_{32}\simeq0.2\pm0.1$ (Walter, Dufour, \&
Hester 1992), which, given the calibration cited (Equation 3), leads
to $12+log(O/H)=8.7\pm0.02$, in excellent agreement with the Prieto,
Lambert, \& Asplund (2001) measurement.}

KP03 show that strong line equivalent width ratios
are tightly correlated with flux ratios and can be used to obtain 
a metallicity-sensitive parameter akin to $R_{23}$ for a wide range of
galaxy colors and emission line properties.  
In this paper we follow the 
prescriptions of KP03.  
Using equivalent widths, equations 5 and 6 become

\begin{equation}
y \equiv\ \log(EWO_{32}) \equiv\ \log\Biggr(
	{{EW_{[O~III]\lambda4959} + EW_{[O~III] \lambda 5007}}
	\over{EW_{[O~II]\lambda3727}}}\Biggr),
\end{equation}

\noindent and

\begin{equation}
x\equiv\ \log({EWR_{23}})\equiv\ \log\Biggr({{EW_{[O~II] \lambda3727} + EW_{[O~III]
\lambda4959} + EW_{[O~III] \lambda 5007}}\over{EW_{H\beta}}}\Biggr).
\end{equation}

\noindent As noted in KP03,
the use of equivalent width ratios, rather than line flux ratios, has
the particular advantage of being less sensitive to the (unknown)
amount of extinction within the host galaxy, at least if the reddening
toward the gas and stars is similar.\footnote{Calzetti, Kinney, \&
Storchi-Bergmann (1994) present evidence that this assumption may be
invalid for some galaxies.}  The range of emission line equivalent widths
and galaxy colors among the DGSS sample studied here falls within the
range where $EWR_{23}$ is a good surrogate for $R_{23}$.  The Appendix
shows histograms of DGSS galaxy colors and equivalent widths compared
to the local galaxy samples used by KP03 to validate the $EWR_{23}$
approach.  Following KP03, we perform a rough correction for stellar
absorption by adding a constant 2 \AA\ to the equivalent widths of
$H\beta$ in Table~1.  This correction has the effect of decreasing $EWR_{23}$ and
increasing the computed oxygen abundance.  The impact of this
correction is negligible in galaxies with large $EW_{H\beta}$ but
raises O/H by as much as 0.08 dex for galaxies with the smallest
$EW_{H\beta}$.

Table~1 records the oxygen abundances, 12+log(O/H), derived from
our tabulated emission-line equivalent widths using the $EWR_{23}$
prescription above.  The galaxies range between 12+log(O/H)=8.4 and
12+log(O/H)=9.0, with typical {\it random} measurement uncertainties of
0.03 to 0.12 dex.  An additional uncertainty of $\sim$0.15 dex in O/H,
representing uncertainties in the photoionization models and
ionization parameter corrections for empirical strong-line calibration,
should be added in quadrature to the tabulated measurement errors.

In order to look for differences between local and DGSS galaxy samples 
that could potentially bias our conclusions,
Figure~\ref{zMBOH} shows the relationship between redshift, $M_B$,
$(B-V)_0$ color, $EW_{H\beta}$, $L_{H\beta}$, and oxygen abundance for
the 64 objects in our sample compared to the three local galaxy samples.
Solid symbols distinguish DGSS objects by redshift: ``low'' (17
objects; $0.26<z<0.40$), ``intermediate'' (21 objects; $0.40<z<0.60$),
and ``high'' (26 objects; $0.60<z<0.82$).  Dots, crosses, and
3-pointed triangles designate the KISS, K92, and NFGS local galaxy
comparison samples.  Here, the nearby comparison samples have been
culled using the same emission-line selection criteria as the DGSS
galaxies.

The redshift--luminosity panel of Figure~\ref{zMBOH} shows a
correlation.  As expected for a flux-limited survey, more high
luminosity objects and fewer low-luminosity objects are detected in
the highest redshift bin compared to the lowest redshift bin. The DGSS
galaxies are bluer than the mean NFGS galaxy but are consistent with the
bluest local galaxies.  The DGSS galaxies also have, on average, larger
emission line equivalent widths than any of the local samples.  
The lower left panel compares the
oxygen abundance versus redshift.   The oxygen abundances of galaxies within each
redshift bin correlate strongly with blue luminosity and less strongly
with $H\beta$ luminosity.  There is no significant correlation between
color and metallicity or between equivalent width and metallicity.
The redshift-metallicity (lower right) panel reveals that the zero point of the L-Z
correlation is displaced toward higher luminosities for larger
redshifts:  DGSS galaxies lie predominantly on the upper envelope of
the local galaxy samples.  While most of the DGSS sample galaxies are
consistent with a luminosity-metallicity trend, the two least luminous
objects, 092-1375 and 172-1242 with $M_B\sim{-16.5}$, lie well off the
correlation if the empirical line-strength to-metallicity conversion
is blindly applied.  Based on studies of other galaxies of similar
luminosity ($M_B\simeq-16.5$; Skillman, Kennicutt, \& Hodge 1989), we
suspect that these two galaxies do not belong on the the upper branch
of the $R_{23}$--O/H relation.  They lie in the
``turn-around'' region between the upper and lower branches where the
strong-line calibration is particularly uncertain.  If these objects
lie on the lower branch of the $R_{23}$ calibration, then they have
metallicities near 12+log(O/H)=8.1.  Unfortunately neither of these
objects have $H\alpha$ or [N~II] detections, so there is no way to way
to discriminate between the two metallicity branches.  Due to their
metallicity ambiguity, we remove these two galaxies from any further
analysis.

\subsection{Comparison with Local Galaxies}

We turn now to a detailed comparison of the luminous and chemical
properties of the remaining DGSS sample to local galaxies and other
intermediate-redshift galaxies. We use subsets of the same three sets
of local galaxies, constrained now also to be brighter than $M_B =
-17$ to better match our sample, and to have $F_{[N II]\lambda 6584} /
F_{H\alpha} > 0.15$ in order to assure they lie on the metal-rich
branch of the $R_{23}$ relation.  The K92, NFGS and KISS samples
yielded 22, 36 and 80 galaxies, respectively, meeting these additional
cuts.

Figure~\ref{LZ} shows the best-fit luminosity-metallicity relation for
DGSS galaxies compared to the three local galaxy samples.  Filled
symbols denote DGSS galaxies in the redshift ranges $z=0.26-0.4$,
$z=0.4-0.6$, and $z=0.6-0.82$ as in Figure~\ref{zMBOH}.  Crosses,
skeletal triangles and dots denote the K92, the NFGS and the KISS
local samples, respectively.  The eight open squares show the
best-measured objects in the $0.6<z<1.0$ field galaxy study by Carollo
\& Lilly (2001; CL01).\footnote{Magnitudes have been converted from
the $H_0=50$, $\Omega_M=1.0$ ($q_0=0.5$) cosmology adopted in that
study.  Oxygen abundances are computed using the published $R_{23}$
and $O_{32}$ values from that work.} Open triangles show the
$z=0.2-0.5$ objects from Kobulnicky \& Zaritsky (1999; KZ99) which
meet the same emission line selection criteria as the DGSS.  Open
stars are the high-redshift ($z>2$) galaxies from Kobulnicky \& Koo
(2000; KK00) and Pettini \etal\ (2001; Pe01).  Lines represent linear
least squares fits to each sample, taking into account the
uncertainties on both O/H and $M_B$.\footnote{To be precise, we used the FITEXY routine
of Numerical Recipes (Press \etal\ 1990)
as implemented in IDL Astronomy Users library. For local galaxy
samples we adopted $1\sigma$ uncertainties of $\delta(M_B)=0.2$ mag and
$\delta(O/H)=0.1$ dex. For the DGSS galaxies we adopt  $\delta(M_B)=0.2$ mag and
we add 0.1 dex in quadrature with the $\delta(O/H)$ listed in Table~1 to 
account the systematic uncertainties plus measurement errors.}  Becuase errors on $M_B$ are usually smaller 
than the errors on O/H, this fit approaches a fit of metallicity on luminosity. 
In the upper left panel of local galaxies the
long-dashed line marks the fit to the combination of all three local
samples (K92+NFGS+KISS) while the heavy dashed line shows the fit to
just the K92+NFGS data.  Formal results from these fits and uncertainties
appear in Table~2.  The local galaxy samples were fit both collectively and
then individually because they exhibit different distributions.
The KISS galaxies lie preferentially to the right (metal-rich and
underluminous) side of the best-fit L-Z relation
for the combined K92+NFGS+KISS sample.  One possible explanation
for this offset is that KISS spectra were obtained with 2-3\arcsec\
fibers or slits centered on the nucleus of each galaxy.  Spectra dominated
by emission from the inner several kpc will preferentially sample
the metal-rich nuclei of galaxies and bias the KISS sample toward
higher metallicities.   For this reason, we use only the fit to the K92+NFGS 
samples (heavy dashed line in Figure~\ref{LZ}) for comparison to the DGSS
galaxies.   

Figure~\ref{LZ} shows a representative error bar in the lower left
panel indicating the typical statistical uncertainties of
$\sigma(M_B)=0.2$ mag and $\sigma(O/H)=0.10$ dex.  It is worth
emphasizing here that the oxygen abundances for all samples
represented in Figure~\ref{LZ} have been computed in an identical
manner from measured emission line equivalent widths.  No corrections
for internal reddening have been made.  We have adopted the
recommendation of KP03 and increased all local and DGSS galaxy
$H\beta$ equivalent widths for 2 \AA\ of stellar Balmer absorption.
The absolute B magnitudes have been computed (or converted from
published values) assuming the $H_0=70$, $\Omega_M=0.3$,
$\Omega_\Lambda=0.7$ cosmology.  K-corrections have been applied to
the high-redshift samples based on multi-band rest-frame B-band and UV
photometry.  The various samples represented should be directly
comparable, although they might not be completely representative of
the entire population of galaxies at that redshift.

Figure~\ref{LZ} reveals a striking evolution in the slope and zero
point of the best fit luminosity-metallicity correlation with
redshift.  This offset between local and DGSS samples is similar even
if other fitting schemes are adopted (e.g., unweighted fits, or O/H
errors only). Although considerable scatter exists within a given
sub-sample, {\it the more distant galaxies are markedly brighter at a
fixed oxygen abundance.}  Alternatively, galaxies of a given
luminosity are systematically more metal-poor at higher redshifts. The
best fit relation for the local K92+NFGS samples is
$12+log(O/H)=(-0.052\pm 0.009)~M_B + 7.74$.  For the $z\sim0.7$ DGSS
galaxies, we find $12+log(O/H)=(-0.128\pm 0.012)~M_B + 6.04$.  Note
that these slopes differ considerably from other published L-Z
relations for local galaxies.  Melbourne \& Salzer (2002), for
example, find $12+log(O/H)=-0.24~M_B + 4.0$.  Some of this difference
is due to the different metallicity calibrations used in that work
compared to our study.  The majority of the difference is due to the
restricted range of galaxy luminosities and metallicities in the
present DGSS sample.  Melbourne \& Salzer (2002) included all KISS
galaxies on both halves of the doubled-valued $R_{23}$ to O/H
relation, down to magnitudes of $M_B=-13$.  They find evidence for a
change in slope of the L-Z relation at low luminosities which alters
the slope of their overall fit compared to that for our more
restricted luminosity range.

More important than the formal zero point differences of the fits are
the luminosity and metallicity offsets of the L-Z relation near the
midpoints of the sample distributions.  These midpoints are
approximately at a metallicity of $12+log(O/H)\sim8.7$ and a magnitude
of $M_B=-20$.  The low-redshift DGSS ($0.2<z<0.4$) subsample is as
much as 1.6 magnitudes brighter and/or 0.10 dex more metal-poor than
the mean of the local sample.  The intermediate- and high-redshift
DGSS subsamples are 2.0 and 2.4 magnitudes brighter than the local
samples of similar luminosity and/or as much as 0.15 dex more
metal-poor.  High-redshift $z\sim3$ galaxies show a larger offset of
at least 3 magnitudes compared to the local sample, although the
uncertainties on O/H are large for those objects.  In the highest
redshift DGSS bin, there is a statistically significant change in the
best-fit slope of the L-Z relation in the sense that the least
luminous galaxies at $0.6<z<0.8$ are more offset from the local L-Z
relation compared to their more luminous counterparts at the same
redshift.

Figure~\ref{LZ} shows that there is considerable dispersion among the
local samples, such that some of the brightest, most metal-rich local
galaxies overlap the $0.6<z<0.8$ L-Z relation.  At luminosities
fainter than $M_B=-19$, however, local galaxies are all 1-3 magnitudes
fainter, for a given metallicity, compared to the $0.6<z<0.8$ objects.
The change in slope of the L-Z relation at $0.6<z<0.8$ is in the sense
that the low-luminosity DGSS galaxies in this redshift interval show
greater offsets from the local samples than the high-luminosity
galaxies.  The lack of DGSS galaxies in the lower right corner of the
plot (low luminosity but high metallicity) represents a real offset
from the local relation and is not due any selection effect we can
identify.  Galaxies that might occupy this locus near $M_B=-19$,
$12+log(O/H)=8.7$ would have measurable emission lines and line
ratios, $R_{23}$, identical to the more luminous galaxies with
$M_B=-21.5$.  Although the ratios of [O~III] and [O~II] line strengths
to H$\beta$ line strengths drop with increasing metallicity, the
sample selection process would not preferentially exclude objects with
low [O~III]/H$\beta$ (high metallicity) in favor of those with high
[O~III]/H$\beta$ (low metallicity).  If such low-luminosity,
high-metallicity objects existed, they would have been detected and
included in the analysis.  In our selection of objects from the DGSS
database, we found only 3 objects (at $z=0.36,z=0.76,z=0.77$) with
measurable $H\beta$ but immeasurably weak [O~III] (S/N$<8:1$).  In all those
cases, poor sky subtraction is clearly the cause of the weak [O~III].
Furthermore, there is no correlation between $EW_{H\beta}$ and O/H
(see Figure~\ref{zMBOH}), as might be expected if selection effects
had preferentially removed objects with weak line strengths from the
sample or if the corrections for Balmer absorption dominated the
errors in a systematic way.  Although other selection effects in the
local or distant samples may give the appearance of
luminosity-metallicity evolution from $z=0.8$ to the present, we
proceed in the next section to show that we can rule out several
obvious selection biases.

\subsection{Examination of Systematics with Color and Equivalent Width}

Selection biases in the local and distant galaxy samples could produce
the signature of an evolving L-Z relation seen in Figure~\ref{LZ}.
For instance, comparing local galaxies having moderate $EW_{H\beta}$
to the rare DGSS galaxies with strong starbursts and
uncommonly large $EW_{H\beta}$ values
could lead to misleading conclusions about the nature of distant
versus local populations.  We investigated sources of 
selection bias by examining the residuals in the L-Z relation as a
function of other fundamental galaxy parameters. Using unweighted
linear least-squares fits to the KISS, NFGS, and DGSS samples
separately, we computed residuals for each sample and searched within 
each sample for correlations with galaxy color, $H\beta$ luminosity, size, and
equivalent width.  Only $EW_{H\beta}$ is marginally correlated with
L-Z residuals, and only for the DGSS galaxies.  

Figure~\ref{LZresidew} plots galaxy $EW_{H\beta}$ versus luminosity
residuals, $\delta{M_B}$, from the above best-fit linear relation for each
subgroup in the L-Z plane.  There is no correlation for the KISS
sample or for the NFGS sample.  We find a small trend for the highest-redshift 
DGSS galaxies which is driven by the handful of
galaxies with extreme values of $EW_{H\beta}$, but the correlation
coefficient indicates an equal or greater correlation could be
achieved in a random sampling of points at least 20\% of the time.
{\it The lack of a significant correlation between magnitude residuals
and $EW_{H\beta}$ is evidence against the hypothesis that extreme
starbursts, which temporarily increase the emission line strength and
overall luminosity, are a source of dispersion in the L-Z relation.}
Hence, the larger emission line equivalent widths in the DGSS samples
compared to the local samples do not bias the distant galaxies toward
larger luminosities and are not responsible for the observed L-Z
relation offsets.  Further evidence for this conclusion comes from
Figure~\ref{LZlim}, which shows only those DGSS and local KISS, NFGS,
and K92 galaxies with the restricted EW criterion $EW_{H\beta}<20$
\AA.  Best fit lines to each sample show that the basic conclusion
remains the same: distant DGSS galaxies are 1-2 magnitudes more
luminous than local galaxies of comparable metallicity.  This result
does not appear to be due to some kind of selection effect since the
DGSS sample analyzed here spans the entire range of luminosities from
the entire DGSS sample in each redshift interval, i.e., we are not
simply selecting the brightest objects from each redshift interval, as
shown by Figure~\ref{select}.

Figure~\ref{select} hints at another kind of selection effect which
may lead to erroneous conclusions regarding differences between DGSS
and local galaxy samples.  All of the DGSS galaxies have very blue
colors with $B-V<0.6$, while only a small fraction of local galaxies
are as blue.  The DGSS samples are likely to contain a larger fraction
of extreme star-forming galaxies with younger light-weighted stellar
populations.  As a second means of examining selection/population
differences between the local and DGSS samples, we have constructed a
third version of Figure~\ref{LZ} using only local galaxies with the
bluest colors.  Figure~\ref{LZlimBOTH} shows only those DGSS and local
K92, KISS, and NFGS galaxies with $B-V<0.6$ {\it and} $EW_{H\beta}<20$
\AA.  Although the number of NFGS and KISS data points are fewer, best
fit lines to each sample and the basic conclusions remain the same.
Another way to look at this result is to say that the $0.6<z<0.8$
galaxies are 10\% to 80\% (0.04 to 0.25 dex, depending on $M_B$) more
metal-poor than local galaxies of comparable luminosity.  Selection
effects such as differences in color distributions or EW distributions
between the local and DGSS samples do not appear to be responsible for
the observed variation of the L-Z relation with epoch.

\subsection{Evolution of the L-Z Relation}

One straightforward interpretation of Figure~\ref{LZ},
Figure~\ref{LZlim}, and Figure~\ref{LZlimBOTH} is to say that emission
line galaxies {\it have undergone considerable luminosity evolution,
$\Delta M_B\simeq1-3$ mag, from $z\sim0.8$ (7 Gyr ago) to the present
and as much as $\Delta M_B\simeq3-4$ mag since $z\sim3$ (11 Gyr ago).}
This amount of brightening is consistent with, or greater than than,
the interpretation of Tully-Fisher evolution studies out to $z\sim1$
(e.g., Forbes, 1995; Vogt \etal\ 1997; Simard \& Pritchet 1998;
Ziegler \etal\ 2002).  An alternative way of understanding our new
result is to say that emission-line field galaxies have experienced {\it a
measurable degree of metal enrichment (up to a factor of 2.0 for the
least luminous objects) at constant luminosity } since $z\sim0.8$.  The
arrows in Figure~\ref{LZlimBOTH} indicate evolution in the L-Z plane
caused by constant star formation, passive evolution, metal-poor gas
inflow, and star formation bursts and/or galaxy mergers.  Constant
star formation drives galaxies mainly to the right in Figure~\ref{LZlimBOTH}
(i.e., metal enrichment at constant luminosity) while passive
evolution moves galaxies downward (i.e., fading at constant
metallicity).  Some combination of these two processes are probably
responsible for evolving the $z=0.6-0.8$ galaxies into the region
occupied by today's $z=0$ galaxies, provided that the evolved
descendants of the DGSS sample can be found among the local samples.
Given that many of the DGSS galaxies appear to be disklike systems
with active star formation, it is reasonable to suppose that a
significant fraction of them are still disklike star-forming systems
at the present epoch.

The apparent evolution in the luminosity-metallicity relation seen in
Figures~\ref{LZ} and \ref{LZlimBOTH} stands in contrast to the
conclusions of Kobulnicky \& Zaritsky (1999) and Carollo \& Lilly
(2001), who found little or no evolution in the redshift ranges
$0.1<z<0.5$ and $0.5<z<1.0$, respectively.  Open squares in
Figure~\ref{LZ} show that the CL01 objects fall at the high-luminosity
end of the luminosity-metallicity correlation where the local and
``distant'' samples overlap.  A comparison of the CL01 objects to the
local samples and to the brightest DGSS galaxies shows that they
have luminosities and metallicities similar to our most luminous objects
and are consistent with the L-Z relation at that epoch.  There is very little
difference between the L-Z relations at $0.6<z<0.8$ and $z=0$ at the
luminous end.  Hence, the
conclusion of CL01, namely that their galaxies lie near the
L-Z relation for local galaxies, is appropriate given the restricted magnitude range
of that sample.  The new conclusion based on Figure~\ref{LZ} is made
possible by the addition of larger numbers of lower-luminosity
galaxies which greatly extend the luminosity and metallicity
baselines.  Although the Kobulnicky \& Zaritsky (1999) sample included
a broad range of luminosities, the least luminous objects were at the
lowest redshifts where offsets from local galaxies are least
pronounced.  Furthermore, Kobulnicky \& Zaritsky (1999) used a local
comparison sample that included only a restricted subset of galaxies
from the Kennicutt (1992a,b) spectral atlas.  Objects from this sample
show increased scatter toward low metallicities at fixed luminosity,
due mostly to the errors introduced by variable, unknown amounts of
stellar Balmer absorption.  The increased scatter in the local sample
masked the trend which has now become apparent in Figure~\ref{LZ}.
The global shift and slope change of the L-Z relation between $z=0$ and $z=0.8$ supports
the conclusions of Kobulnicky \& Koo (2000), Pettini
\etal\ (2001) and Mehlert \etal\ (2002) at higher redshifts, who found 
galaxies to be overly luminous for their metallicity at $2.2<z<3.4$.

How do our results compare to conclusions about the evolution of the
luminosity function (LF)?  Lin \etal\ (1999) use early CNOC2 data on
$>2000$ galaxies over the redshift range $0.12<z<0.55$ to conclude
that, among late-type galaxies (the ones most comparable to the DGSS
objects selected here), changes in the LF are most consistent with
{\it density} evolution, and little or no {\it luminosity evolution}.
However, Cohen (2002), based on studies in the Hubble Deep Field North
vicinity over the redshift range $0.01<z<1.5$, concluded that
emission-line objects show {\it moderate luminosity evolution} and
{\it little density evolution}.  Interpreting Figure~5 of Lin \etal\
(1999) as luminosity evolution suggests that, between $z\simeq0.70$
and $z\simeq0.20$, late-type galaxies with $M_B=-19$ fade by $\simeq$1
mag, while galaxies with $M_B=-21$ fade by $\simeq0.2$ mag.  
More recent results of Wolf \etal\ (2002; COMBO-17 survey) 
show a decrease in the luminosity of $M^*$ galaxies from z$\sim1$
to the present by as much as 2 magnitudes, dpending on spectral type.
Given the uncertainties in the distant luminosity functions, this amount of
fading is consistent with the amount of fading
required for the $0.6<z<0.8$ DGSS galaxies to match the local L-Z relation
in Figure~\ref{LZ}.  However, this consistency does not prove the fading model,
and other possibilities are discussed below.

\section{Discussion}
\subsection{Simple Galaxy Chemical Evolution Models }

The addition of chemical information on galaxies at earlier epochs
provides a new type of constraint on theories of galaxy formation and
evolution.  If local effects such as the gravitational potential and
``feedback'' from supernova-driven winds are the dominant regulatory
mechanisms for star formation and chemical enrichment, then the L-Z
relation might be nearly independent of cosmic epoch.  The
semi-analytic models of Kauffman (1996) and Somerville \& Primack
(1999) show little or no evolution in the L-Z relation with epoch,
although uncertainties in the model-dependent prescriptions for
stellar feedback and galactic winds may have large impacts on the
computed chemical and luminous properties.  Based on current
understanding that the volume-averaged star formation rate was higher
in the past (Pei \& Fall 1995; Madau 1996), and that the overall
metallicity in the universe at earlier times was correspondingly
lower, we might expect galaxies to be both brighter and more metal-poor
in the past.  A high- or intermediate-redshift galaxy sample ought,
therefore, to be systematically displaced from the local sample in the
luminosity-metallicity (L-Z) plane {\it if individual galaxies reflect
these cosmic evolution processes}. The key questions are two-fold:
which dominates---the chemical evolution or luminosity evolution, and
how do those rates differ, if they do, in low-luminosity galaxies
versus high-luminosity galaxies?  Comparison of the data from
Figure~\ref{LZ} to some simple galaxy evolution models can help
distinguish among these possibilities.

In a simple evolutionary model, galaxies begin as parcels of gas which
form stars and produce metals as the gas fraction decreases from 100\%
to 0\%.  The B-band luminosity\footnote{Here, we consider only the
B-band luminosity-metallicity relation.  Since young stars dominate
the B-band luminosity, we expect that much of the scatter of the
B-band L-Z relation may be due to extinction and fluctuations in the star
formation rate.  Ideally, a rest-frame I-band or K-band
luminosity-metallicity relation would be a more sensitive tool for
evolution studies, since it will be less sensitive to these effects
and eliminate the need for the color correction discussed in the
previous section.} is related to the star formation history, which
may vary as a function of time but is generally taken to be
proportional to the mass or surface density of the remaining gas
(e.g., Kennicutt 1998).  For a galaxy that evolves as a ``closed box'',
converting gas to stars with a fixed initial mass function and
chemical yield, the metallicity is determined by a single parameter:
the gas mass fraction, $\mu=M_{gas}/(M_{gas}+M_{stars})$.  The
metallicity, $Z$, is the ratio of mass in elements heavier than He to
the total mass and is given by

\begin{equation}
Z= Y ln(1/\mu) \label{mu},
\end{equation}

\noindent where $Y$ is the ``yield'' as a mass fraction.  A typical
total metal yield for a Salpeter IMF integrated over 0.2--100
$M_\odot$ is $Y=0.012$ by mass (i.e., 2/3 the solar metallicity of
0.018; see Pagel 1997, Chapter 8). A total oxygen yield for the
same IMF would be $Y_{O}=0.006$.  Effective yields in many local
galaxies range from solar to factors of several lower (Kennicutt \&
Skillman 2001; Garnett 2002), implying either that the nucleosynthesis
prescriptions (e.g., Woosley \& Weaver 1995) are not sufficiently precise, 
that assumptions about the form of the initial mass function are
incorrect, or that metal loss is a significant factor in the evolution
of galaxies.  Garnett (2002) finds that oxygen yields of $Y_{O}=0.001$
to $Y_{O}=0.014$ among local irregular and spiral galaxies are
correlated with galaxy mass, suggesting an increasing amount of metal
loss among less massive galaxies.  High-velocity winds capable of
producing mass loss are observed in local starburst galaxies (e.g.,
Heckman \etal\ 2000) and in high-redshift Lyman break galaxies
(Pettini \etal\ 2001), but the actual amount of mass ejected from
galaxies is difficult to estimate.  Simulations indicate that winds
may be metal-enriched, such that metals are lost from galaxies more
easily than gas of ambient composition (Vader 1987; De Young \&
Gallagher 1990; MacLow \& Ferrara 1999).  Recent Chandra observations
of a local dwarf starburst galaxy, NGC~1569, provide the first direct
evidence for metal-enhanced outflows (Martin, Kobulnicky, \& Heckman
2002). Thus, the closed-box models are probably not appropriate, and
more realistic models including selective metal loss are required.

\subsection{Comparison to \textsc{P\'egase2} Models }

In order to compare the luminous and chemical evolution of galaxies in
the L-Z plane to theoretical expectations, we ran a series of
\textsc{P\'egase2} (Fioc \& Rocca-$\!$Volmerange 1999) models. 
\textsc{P\'egase2} is a galaxy evolution code
which allows the user to specify a range of input parameters. The critical
input parameters include:

\begin{itemize}

\item{$M_{tot}$: The total mass of gas available to form the galaxy.
We explored a range of masses, and we present two representative
masses of $10^{10}$ \mo\ and $10^{11}$ \mo.  No dark matter is
included. }

\item{$\tau_{infall}$: The timescale on which the galaxy is assembled.  
We assume that galaxies are built by continuous infall of
primordial-composition gas with an infall rate that declines over an
exponential timescale, $\tau_{infall}$, as implemented in
\textsc{P\'egase2},

\begin{equation}
M(t) = M_{tot} { {e^{-t/\tau_{infall} } \over{\tau_{infall}} }}.
\end{equation}

\noindent 
We chose two representative gas infall timescales of $\tau_{infall}=1$ Gyr and 
$\tau_{infall}=$5 Gyr.  }

\item{$\alpha(IMF)$: The form of the stellar initial mass function.
We adopt the Salpeter value, $\alpha=-2.35$, between 0.1 and 120 \mo\
(following Baldry \etal\ 2002). }

\item{$Y$: The chemical yields from nucleosynthesis.    
We assign Woosley \& Weaver 1995 B-series models for massive
stars. The effective metal yield of these models is $Y=0.016$.  The
\textsc{P\'egase2} models treat a galaxy as a single zone with uniform
chemical composition.  This oversimplification is adequate to achieve
a basic understanding of how fundamental parameters like star
formation rate and gas supply affect gross galaxy properties like
luminosity and metallicity, but it is insufficient for a detailed
analysis of large galaxies with a range of internal metallicities.  }

\item{$F(SFR)$: The form of the star formation rate. 
We adopt a physically motivated prescription where the star formation
rate is proportional to the mass of accumulated gas (or the gas
density; Schmidt 1959; Kennicutt 1998). }

\item{$A_V$: Extinction due to dust.  We
included an inclination-averaged extinction prescription as
implemented in \textsc{P\'egase2}, but extinction changes the model B
magnitudes by only 0.2 mag.  Extinction is of minor importance in the
present comparison.  }

\item{$T_0$: The time after the Big Bang that a galaxy begins to
assemble.  While not explicitly part of the \textsc{P\'egase2} models,
this parameter is implicitly required to match the ages of
\textsc{P\'egase2} galaxies to a particular redshift and lookback time
for our adopted cosmology.  We
initially assume that galaxies begin to form 1 Gyr after the Big Bang
so that local $z=0$ galaxies have approximate ages of 12 Gyr (in a 13.5 
Gyr old universe) and
DGSS galaxies in our high-redshift bin at $0.6<z<0.8$ have ages of
5.7--6.5 Gyr (6.8--5.8 Gyr lookback times).  For convenience, Table~3 lists the
lookback time and age of the universe at a given redshift for our
adopted cosmology.  We show later that
allowing a range of $T_0$ provides a natural explanation for some of
the features seen in the L-Z relations.  }

\end{itemize}

Our goal in the \textsc{P\'egase2} modeling was to reproduce the
qualitative behavior of Figure~\ref{LZ} where, 1) galaxies at
$0.2<z<0.8$ are displaced toward brighter magnitudes and/or lower
metallicities relative to their $z=0$ counterparts, and 2) DGSS
galaxies at $0.6<z<0.8$ exhibit a change in slope compared to more
local L-Z relations. A significant constraint on the models is the
position of $z\simeq3$ Lyman break galaxies, which lie 2-4 magnitudes
brighter (0.5 dex more metal-poor) than the local L-Z relation.  If
the present-day emission-line samples are at all representative of the
evolved descendants of the $0.6<z<0.8$ and $z\simeq3$
galaxies,\footnote{Most of the DGSS objects studied here have optical
morphologies (see Figure~\ref{SpecIm1}) that qualitatively resemble
disks, and their bulge-to-disk ratios are $B/T << 1$.  Even if the
current levels of star formation decline, their evolutionary
descendants are likely to be systems with substantial disk components.
Thus, it is plausible to think that the descendants of the $0.6<z<0.8$
DGSS galaxies are disk-like star-forming galaxies today, similar to
the K92, KISS, and NFGS objects plotted in Figure~\ref{LZ}.The evolutionary 
relationship between
DGSS and Lyman Break galaxies is less clear.} then a
successful model must reproduce the rapid rise in metallicity and
luminosity in the first 2-3 Gyr (as evidenced by the Lyman break
galaxies), and it must reproduce the lack of significant L-Z evolution
during the past 2-3 Gyr (as evidenced by the similarity of the $z=0.2$
galaxies to the present-day L-Z relation).  This means that the rate of
evolution of the L-Z relation must be rapid during the first few Gyr
of a galaxy's formation, and it must then drop dramatically as a galaxy
approaches the present-day L-Z relation.  Furthermore, large luminous galaxies
should slow their L-Z evolution more than less luminous galaxies.
A slowing of the luminous and/or
chemical evolution of a galaxy must physically correspond either 1) to
cessation of star formation and ultimately to depletion of the gas
available to fuel star formation, or 2) to continued inflow of
metal-poor gas which fuels star formation and dilutes the composition
of the ISM so that a galaxy remains fixed in L-Z space.

Figure~\ref{LZevp2} shows evolutionary tracks of galaxies in the L-Z
plane based on the infall \textsc{P\'egase2} models.  The warm ISM metallicity
by mass, $Z$, is marked along the upper $x$-axis along with the
corresponding closed-box gas mass fraction given by the models
(equivalent to the gas mass fractions predicted by Equation~\ref{mu}
using a total yield of $Y$=0.016) as a comparison.  Pentagons in
Figure~\ref{LZevp2} mark the model galaxies at ages of 1, 2, 4, 8, and
12 Gyr (from left to right).  The numbers in parentheses next to each model point
show the lookback time in Gyr assuming that the galaxy begins its assembly 
1 Gyr after the Big Bang.  

Dashed lines denote models with $\tau_{infall}=1$ Gyr and solid lines
denote models with $\tau_{infall}=5$ Gyr.  The upper set of curves
shows galaxies with $M_{tot}=10^{11}$ \mo\ while the lower curves show
galaxies with $M_{tot}=10^{10}$ \mo (not including dark matter).  
Solid lines mark the best-fit
L-Z relations from Figure~\ref{LZ}.  These models produce the rapid
rise in metallicity in the first few Gyr.  These model masses are in
reasonable agreement with observational estimates which place lower
limits on the {\it dynamical} masses of $z\simeq3$ Lyman break
galaxies at $few\times10^{10}$ \mo\ (Kobulnicky
\& Koo 2000; Pettini \etal\ 2001\footnote{But see 
Somerville, Primack \& Faber 2001 who argue that the stellar masses of
Lyman break galaxies are only few$\times10^9$ \mo.} and indicate star
formation rates of $\sim$50 \mo/$yr^{-1}$. However, these models
over-enrich the ISM at late times after 8 Gyr compared to the data.
They predict gas mass fractions that are as low as 1\% and
metallicities as high as 3 times solar (Z=0.054) after 12 Gyr, at odds
with the observed properties of nearby galaxy samples.

We can achieve better agreement between the models and data by varying
several crucial model parameters.

\begin{itemize} 

\item{{\it Varying the SFR}: Lowering the model star formation rates would
delay the chemical enrichment until later times and lower the
luminosity at any given time.  Effectively, this change produces a
wholesale shift in the model curves to the left and downward in
Figure~\ref{LZevp2}. However, lowering the SFR would also lower the
predicted emission line equivalent widths which may already be too low
in the models compared to the data (mean of $EW_{H\beta}=10$ \AA\
after 8 Gyr versus 15-20 \AA\ in the data).  Thus, changing the SFR
does not improve the model agreement with the data.  }

\item{{\it Varying the chemical yield, $Y$}: As an experiment to
produce models that better resemble the data, we introduced a metal
yield efficiency parameter, $\eta_Z$, which reduces the effective yield
of the models.  Reducing the chemical yields simply moves the model
curves {\it en masse } to the left without changing the spacing of the
model points (i.e., without changing the $\Delta(O/H)$ over a fixed
time interval).  Physically, $\eta_Z$ may represent the fraction of
metals retained in galaxies during episodes of wind-driven metal loss,
or it may represent a correction to the theoretical yields from
stellar populations.  Figure~\ref{LZevp} shows the same models and
data as Figure~\ref{LZevp2}, except that the metal yield has been
reduced to 1/2 ($Y_O=0.003$ for $\eta_Z=0.5$).\footnote{This factor is
not incorporated into the
\textsc{P\'egase2} models in any self-consistent manner.  Here we have
simply reduced the final metallicity of the gas at each timestep by
the factor $\eta_Z=0.5$. }  The $M_{tot}=10^{11}$ \mo\ model with
$\tau_{infall}=1$ Gyr is now in better agreement with the $0.6<z<0.8$
and $z=0$ data at ages of 8-12 Gyr. However, the same model for
$M_{tot}=10^{10}$ \mo\ becomes too metal-rich at ages of 8-12 Gyr
compared to the $z=0$ data.  }

\item{{\it Varying $\tau_{infall}$}: Increasing the infall timescale from
1 Gyr to 5 Gyr, as shown in Figure~\ref{LZevp}, alters a galaxy's
evolution in several respects.  It significantly decreases the
luminosity during the first few Gyr, while producing a slight drop in
the gas-phase metallicity at each timestep.  Longer $\tau_{infall}$ also decreases
the change in metallicity in each timestep, i.e., decreases the
$\Delta(O/H)$ over a fixed time interval, most noticeably at late
epochs.  In the context of the cold dark matter paradigm, the
objects with the largest initial overdensities form most rapidly (e.g.,
Blumenthal \etal\ 1984; Navarro, Frenk, \& White, 1995).  Whether
these first objects grow to become massive galaxies depends on the
environment in which they are conceived.  Initial galaxy cores which
collapse in the vicinity of larger concentrations of baryonic matter
grow to become the massive galaxies at the centers of today's galaxy
clusters.  Objects that form in regions of lower cosmological
overdensity would accrete gas on longer timescales and, if isolated
from external influences, would proceed more slowly through the star
formation and chemical evolution process.  A longer formation
timescale for lower mass field galaxies observed in the DGSS at
$0.6<z<0.8$ can plausibly explain their observed evolution in the L-Z
relation. Figure~\ref{LZevp} demonstrates that $\tau_{infall}=5$ Gyr
provides better agreement than the $\tau_{infall}=1$ Gyr to the
$0.6<z<0.8$ and $z=0$ galaxies.  However, we stress that a single set of model
parameters cannot reproduce either the L-Z relation itself at {\it
any} epoch or the variation in slope of the L-Z relation with epoch.
The L-Z relation predicted by a single set of model parameters at any
epoch would be a vertical line in Figure~\ref{LZevp}!  }

\item{{\it Varying $T_{0}$}: Delaying the epoch at which model
galaxies begin to assemble simply relabels the ages of models.  For
example, setting $T_0=2$ Gyr instead of $T_0=1$ Gyr renumbers the ages
of the model points in Figure~\ref{LZevp} from 1,2,4,8,12 Gyr to
0,1,3,7,11 Gyr. For a given lookback time, the galaxies are younger.}

\end{itemize}

The effects of decreasing $Y$, increasing $\tau_{infall}$, and
increasing $T_0$ are, to some degree, degenerate.  They all serve to
move model points toward the left (metal-poor) side of
Figure~\ref{LZevp}, and we cannot, here, discriminate between their
relative importance. However, within the context of these simple
models, {\it no single parameter can solely
explain both the  slope of the L-Z relation and the change
in slope with lookback time}.  These observed qualities can only be
produced in the context of the \textsc{P\'egase2} models if at least {\it two} of the following conditions 
obtain: 1)
the chemical yield of low-luminosity galaxies are reduced relative to
high-luminosity galaxies (e.g., Garnett 2002), 2) $\tau_{infall}$ is
larger for low-luminosity galaxies relative to high-luminosity galaxies,
3) $T_{0}$ is larger for low-luminosity galaxies relative to
high-luminosity galaxies (e.g., as in the delayed formation scenarios for
dwarfs, Babul \& Rees 1992).\footnote{Some galaxies, particularly
dwarfs, may never exhaust their gas by the normal star formation
process and may never reach high metallicities if the gas reservoir is
cut off or removed prematurely by ram pressure stripping in a cluster
environment or by galactic winds.} Such a scenario, whereby galaxies 
assembly timescales or formation epochs vary
with environment, is similar to that proposed by Sandage, Freeman, \&
Stokes (1970) to explain the differences along the Hubble sequence.

There is considerable observational support for the idea that lossy
galactic winds remove metals and reduce the effective chemical yields
from star formation (e.g., Martin, Kobulnicky \& Heckman 2002; Garnett
2002).  Winds are more effective in low-mass galaxies with lower
gravitational potentials. The overall slope of the L-Z relation may plausibly
be explained by a decreasing effective yield in low-luminosity
(low-mass) galaxies.  Some combination of longer $\tau_{infall}$, and/or
later $T_0$ for low-mass galaxies may then be invoked to explain the
change in slope of the L-Z relation which preferentially displaces
$0.6<z<0.8$ low-luminosity galaxies more dramatically from the local L-Z relation.

Figure~\ref{LZevp4} uses the same \textsc{P\'egase2} models to explore
how one possible combination of reduced metal yields in low-mass
galaxies relative to larger galaxies combined with a later formation
epoch for low-luminosity galaxies, can come close to reproducing both the slope of the
L-Z relation {\it and} its zero point and slope variations with
redshift.  Solid lines show the 5 Gyr infall models from previous
figures.  As in Figure~\ref{LZevp}, we use $\eta_Z=0.5$ for the
$10^{11}$ \mo\ galaxy but here we choose $\eta_Z=0.4$ for the $10^{10}$
\mo\ galaxy.  This variation in yield is well within the range found
by Garnett (2002).  For the $10^{11}$ \mo\ galaxy we adopt a formation
epoch $T_0=1$ Gyr while for the $10^{10}$ \mo\ galaxy we choose a
formation epoch $T_0=3$ Gyr.  The numbers next to each model point
denote the lookback time and the numbers in parentheses denote galaxy age.
Note how the slope of the L-Z relation is in good agreement with a
line connecting the model points at the present epoch, but it is in good
rather poor with the slope of a line connecting the model points at a
lookback time of 7 Gyr (roughly $0.6<z<0.8$).  Moreover, the
DGSS galaxies at $0.6<z<0.8$ are, as a whole, more metal-rich than the model points
at 7 Gyr lookback times, indicating that these particular
\textsc{P\'egase2} models fail to reproduce (i.e., they overpredict)
the magnitude of zero-point evolution in the L-Z relation.
Nevertheless, this combination of later formation epoch and 
reduced yield for low-mass galaxies does a reasonable job of reproducing the
sense of the zero point and slope evolution in the L-Z plane with
redshift, especialyl for the low-luminosity galaxies.  The largest 
discrepancies are for luminous galaxies whose gas-phase metallicity evolves less than
the  \textsc{P\'egase2} models predict.

\subsection{Limitations of the Single-Zone \textsc{P\'egase2} Models }

As noted, the single-zone \textsc{P\'egase2} models in the above figures fail to
reproduce one important feature of the L-Z relation data, namely, the observed
lack of evolution for luminous galaxies between $0.6<z<0.8$ and 
$z=0$.  All of the models in Figures~\ref{LZevp2},\ref{LZevp}, and
\ref{LZevp4} predict an increase of 0.2 -- 0.4 dex in O/H between
lookback times of $\sim7$ Gyr and the present.  Yet, the linear fits
to the local and the $0.6<z<0.8$ DGSS data indicate get little ($<0.1$
dex) change in O/H for the most luminous galaxies ($M_B<-21$).  The
only ways to decrease the $\Delta (O/H)$ per timestep withinthe framework of these
single-zone models are 1)
to drastically reduce the star formation rate, i.e., to slow the
conversion of gas into stars and heavy elements, or 2) to increase the
infall timescale beyond the 5 Gyr models.  However, such a decrease in star formation
rate or increase in $\tau_{infall}$
would produce model galaxies which are dramatically more metal-poor
and gas-rich at a given age compared to the present models, and would
be in poor agreement between the models and the data.  We can find no
reasonable range of \textsc{P\'egase2} model parameters which
reproduce the similarity of luminous local galaxies to $0.6<z<0.8$
luminous DGSS galaxies.

The resolution to this crisis may lie in the
considerable internal complexity of large spiral galaxies (e.g., radial
chemical gradients, spatially variable extinction and star formation
rates), which single-zone model like \textsc{P\'egase2} do not include.
More
sophisticated models which treat galaxies as a collection of annuli
with differing gaseous and chemical properties, for example, might
more accurately reproduce the observed characteristics of the large
galaxies seen locally and in the DGSS.  The lack of L-Z evolution
among luminous galaxies in Figure~\ref{LZevp} might be explained if
star formation and chemical enrichment in these objects proceeds
radially from the inside-out.  If the star formation in local galaxies
is dominated by activity at a larger radius than the star formation
activity in $0.6<z<0.8$ DGSS galaxies, then the luminosity-weighted
DGSS spectra will preferentially reflect the chemical composition at 
different (i.e., smaller) galactocentric radii in DGSS galaxies than in local
galaxies.  Although local luminous spiral galaxies may, on whole, be
substantially more metal-rich today than at $0.6<z<0.8$, their
emission line luminosities today would be dominated by star formation in
comparably-enriched regions, but at larger radii, compared to the DGSS
galaxies.  Observations of the mean star formation radius within disk
galaxies as a function of redshift could test this hypothesis.  
Regardless of whether this particular hypothesis is true,
additional information about the spatial distribution of star forming
regions in distant objects (e.g., using the Hubble Space Telescope), will
lead to a greater understanding of the luminous and chemical histories of
large galaxies.  

One additional free parameter not incorporated into the
\textsc{P\'egase2} models is the role of starbursts on timescales of a few Myr 
that occur in galaxies that evolve on long (Gyr) timescales according
to the model prescriptions above. Studies of local galaxies show that
their star formation histories are not smooth but instead are
punctuated by periods of enhanced activity.  Such short-term
enhancements in the star formation rate of galaxies otherwise evolving
smoothly can elevate the luminosity and emission line equivalent
widths by factors of several.  The observed emission line equivalent
widths in DGSS galaxies average 10-20 \AA, but some exceed 50!  These
characteristics would be better fit by superimposing a burst of star
formation of $\sim10^{6-7}$ \mo\ on the model galaxies.  The impact of
instantaneous bursts on the evolution in the L-Z plane would be to
create spikes along the model tracks toward higher luminosities.  The
amplitude of the spikes would scale with the amplitude of the burst,
and could conceivably be up to several magnitudes and would persist up to a few
100 Myr. Given the morphological evidence for ongoing mergers among
the DGSS sample, evidenced by inspection of Figure~1, it seems possible
that short-term enhancements in the star formation rate are
responsible for the more extreme luminosities and emission line
equivalent widths among the sample.  Examination of the
\textsc{P\'egase2} models used here shows that as the $EW_{H\beta}$
declines from 50 \AA\ to 10 \AA, the B-band luminosity drops by 0.7
mag. If we were to try to correct the magnitudes of the DGSS samples
by some factor to account for their larger equivalent widths compared
to the local samples, this factor would be less than 0.7 mag, not
enough to account fully for the larger 1-2 mag offsets observed in the
L-Z relations.  This is further evidence against the possibility that
the offset of the L-Z relation at $z=0.7$ is merely caused by
starbursts with temporarily elevated luminosities and equivalent
widths, although this may be part of the answer.

\section{Conclusions}

Observations of star-forming galaxies from the Deep Extragalactic
Evolutionary Probe (DEEP) survey of Groth Strip galaxies in the
redshift range $0.26<z<0.82$ show a correlation between B-band
luminosity and ISM oxygen abundance, like galaxies in the local
universe.  Both the slope and zero-point of the L-Z relation evolve
with redshift.  DGSS galaxies are 1---3 mag more luminous at
$0.6<z<0.8$ than $z=0$ galaxies of similar metallicity.  Said another
way, galaxies of comparable luminosity are 0.1-0.3 dex more metal-rich
at $z=0$ compared to $0.6<z<0.8$.  The change in slope of the L-Z
relation is most pronounced for DGSS galaxies at $0.6<z<0.8$ and
indicates that the apparent chemo-luminous evolution is more dramatic
for the least luminous galaxies.  We have ruled out several selection
effects and biases in the galaxy samples which could produce this
result.  Evolution of galaxies in the L-Z plane over the redshift
range $z=3$ to $z=0$ appears to result from a combination of fading
and chemical enrichment for plausible single-zone galaxy evolution
models.  Within the context of these models, the slope of the L-Z
relation and the observed evolution of {\it slope and zero-point} with
redshift imply that that at least two of the following are true for
star-forming galaxies out to $z\simeq0.8$: 1) low-luminosity galaxies
have lower effective chemical yields than massive galaxies, 2)
low-luminosity galaxies assemble on longer timescales than massive
galaxies, 3) low-mass galaxies began the assembly process at a later
epoch than massive galaxies.  Comparison of observed ISM metallicities
to simple galaxy evolution models suggests that loss of up to 50\% of
the oxygen is necessary to avoid over-enriching the galaxies at the
observed redshifts.  While the models do a reasonable job of
reproducing the evolution of the less luminous galaxies in our sample
($M_B\sim -19$), the models do a poor job of reproducing the relative
lack of chemo-luminous evolution seen among the luminous galaxies in
our sample ($M_B\sim -22$), suggesting a breakdown of the single-zone
approximation.  We predict that DGSS galaxies have significantly
higher (20\%-60\%) gas mass fractions than comparably luminous local
galaxies.  Such a systematic difference in the mean gas mass fractions of
distant galaxies could, in principle, be observed with instruments such as the
proposed radio-wave Square Kilometer Array.

\acknowledgments

We thank John Salzer for insightful conversations and for providing
the KISS data in electronic form, Shiela Kannappan for a helpful
discussion about the NFGS, Matt Bershady and James Larkin for
scientific inspiration.  Detailed suggestions by a very perceptive
referee greatly improved this manuscript.  H.~A.~K was supported by
NASA through grant
\#HF-01090.01-97A awarded by the Space Telescope Science Institute
which is operated by the Association of Universities for Research in
Astronomy, Inc. for NASA under contract NAS 5-26555 and by NASA
through NRA-00-01-LTSA-052.  This work was also made possible by NSF
grants AST95-29098 and AST00-71198 and NASA/HST grants AR-07532.01,
AR-06402.01, and AR-05801.01.

\appendix

\section{Applicability of $EWR_{23}$ Method to DGSS Galaxies  }

Kobulnicky \& Phillips (2003) showed that the quantity $EWR_{23}$ is a
good surrogate for the metallicity indicator $R_{23}$ over a wide
range of galaxy colors and emission line properties.  However, the
correlation between $EWR_{23}$ and $R_{23}$ is tightest for galaxies
with strong emission lines (e.g., $EW_{H\beta}>10$).  Furthermore, the
residuals from the 1:1 correspondence show a slight correlation with
galaxy color, the residuals being smallest for the bluest galaxies.
Figure~\ref{histcompare} reproduces part of Figure~5 from KP03 and
shows the distribution of $EWR_{23}$, $\log~EW_{H\beta}$,
$\log~EW_{[O~II]}$, $\log~(EW_{[O~III]}/EW_{[O~II]})$, and $B-V$,
(histograms) of 66 DGSS galaxies compared to the residuals from the
$EWR_{23}$ vs.  $R_{23}$ relation.  This figure demonstrates that the
DGSS galaxies studied here lie in regimes where the $EWR_{23}$
vs. $R_{23}$ correlation is strong and well-behaved.  For these
galaxies, $EWR_{23}$ is a good surrogate for $R_{23}$.

\clearpage

\begin{figure}
\figcaption[SpecIm1.ps] {Unfluxed spectra for 66 emission-line galaxies culled
from the DEEP survey of the Groth Strip using the criteria discussed in
Section~2, along with their HST F814W images.  Y-axis is counts in ADU where
ADU=$e^- /2.4$.
Objects are ordered by
redshift, as in Table~1, with the 10 possible or probable AGN
at the end.  Markings identify major emission lines of the [O~II]$\lambda$3726/29
doublet, [Ne III] $\lambda3868$, H$\beta$, [O~III]$\lambda\lambda$4959,5007, 
and, where applicable, $H\alpha$.  Other
positive features are generally residuals from night sky lines which
could not be completely subtracted.  Each panel shows
two different spectra, obtained with ``red'' and ``blue'' grating settings.
Seven-digit strings in the upper left 
give the DGSS identification number.  Rectangles on 
the images show spectroscopic slit positions.   (All 10 panels appear 
in the electronic edition only).
\label{SpecIm1} }
\end{figure}

\plotone{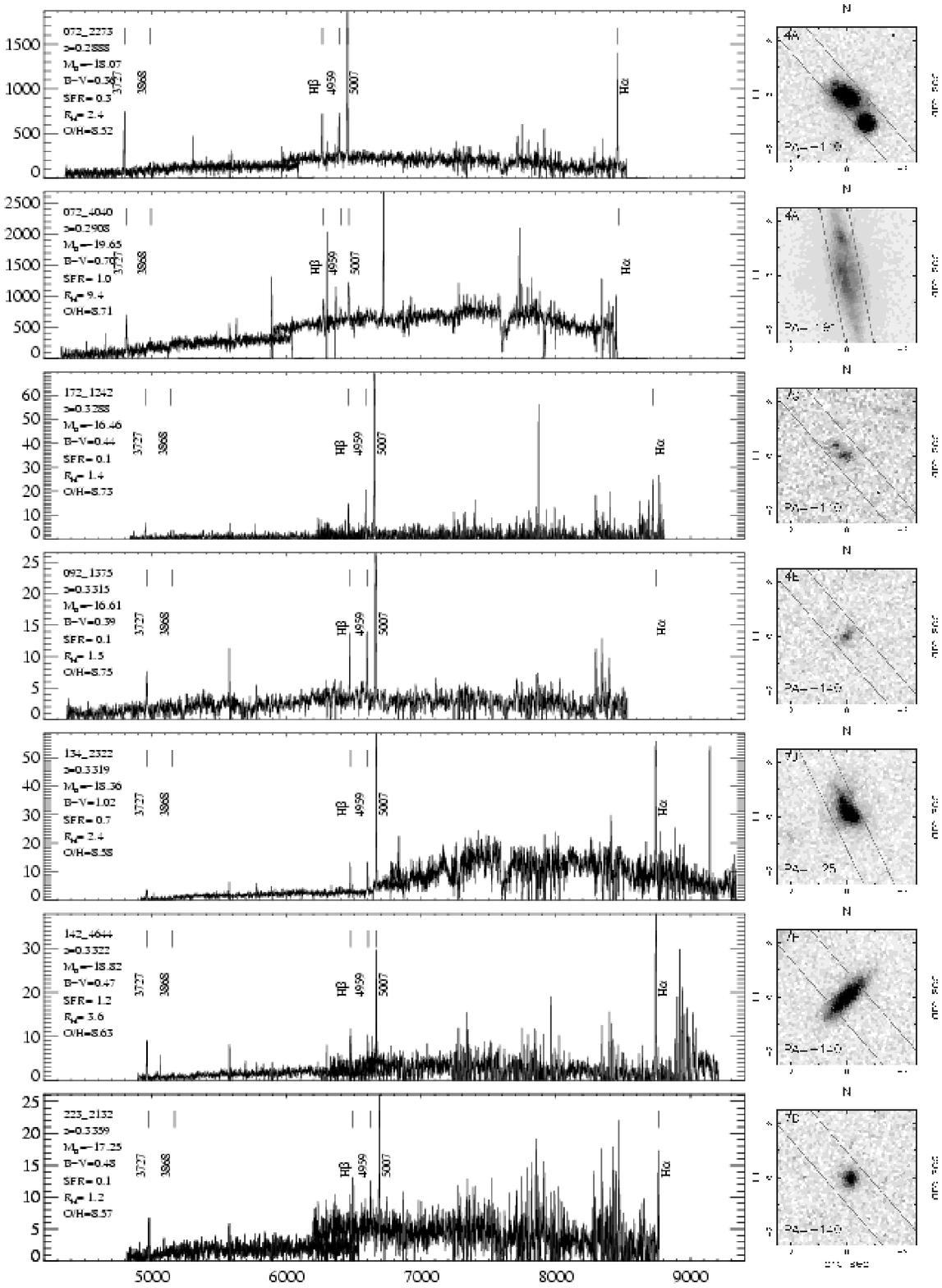}
\plotone{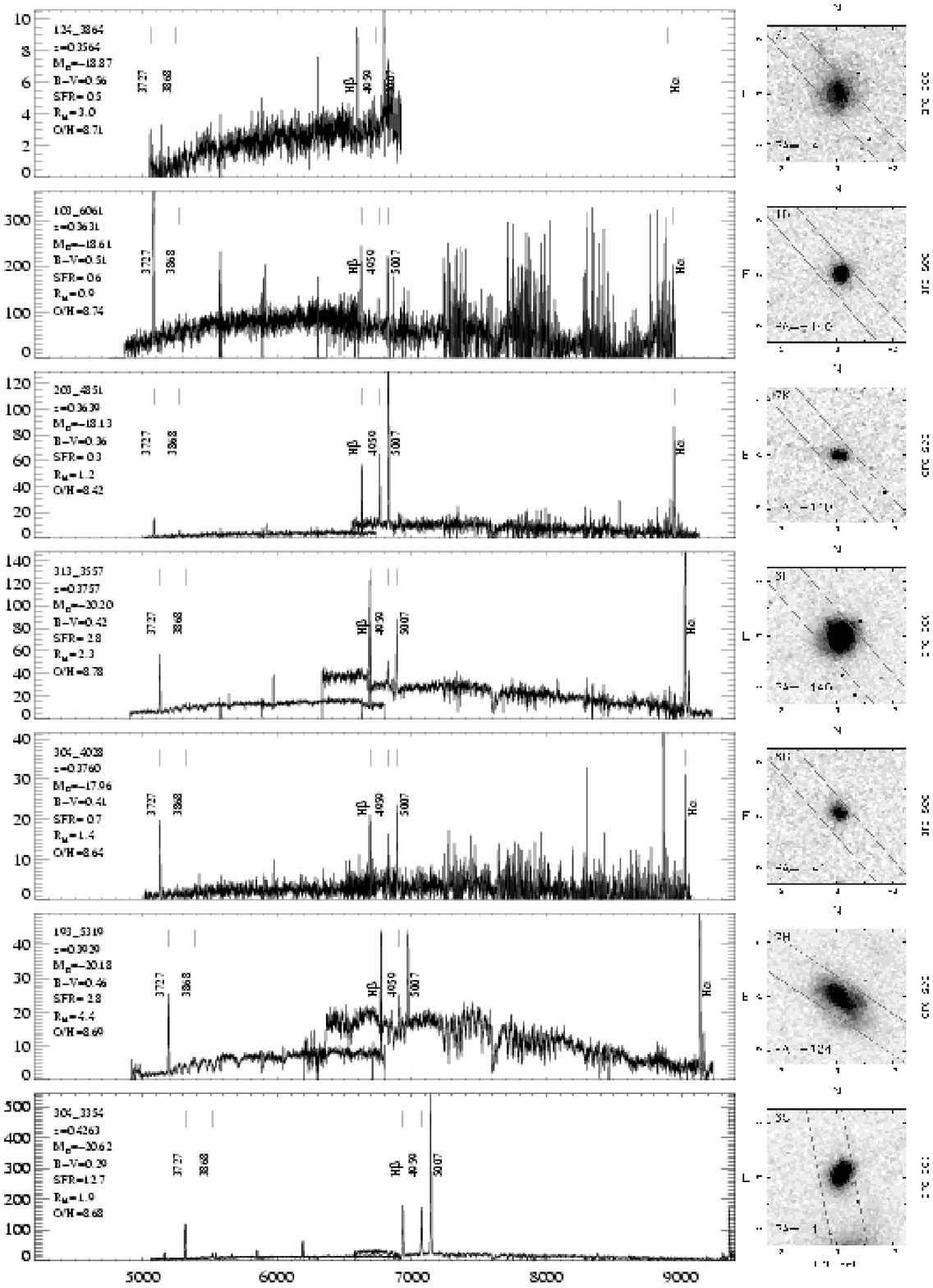}

\clearpage

\begin{figure}
\plotone{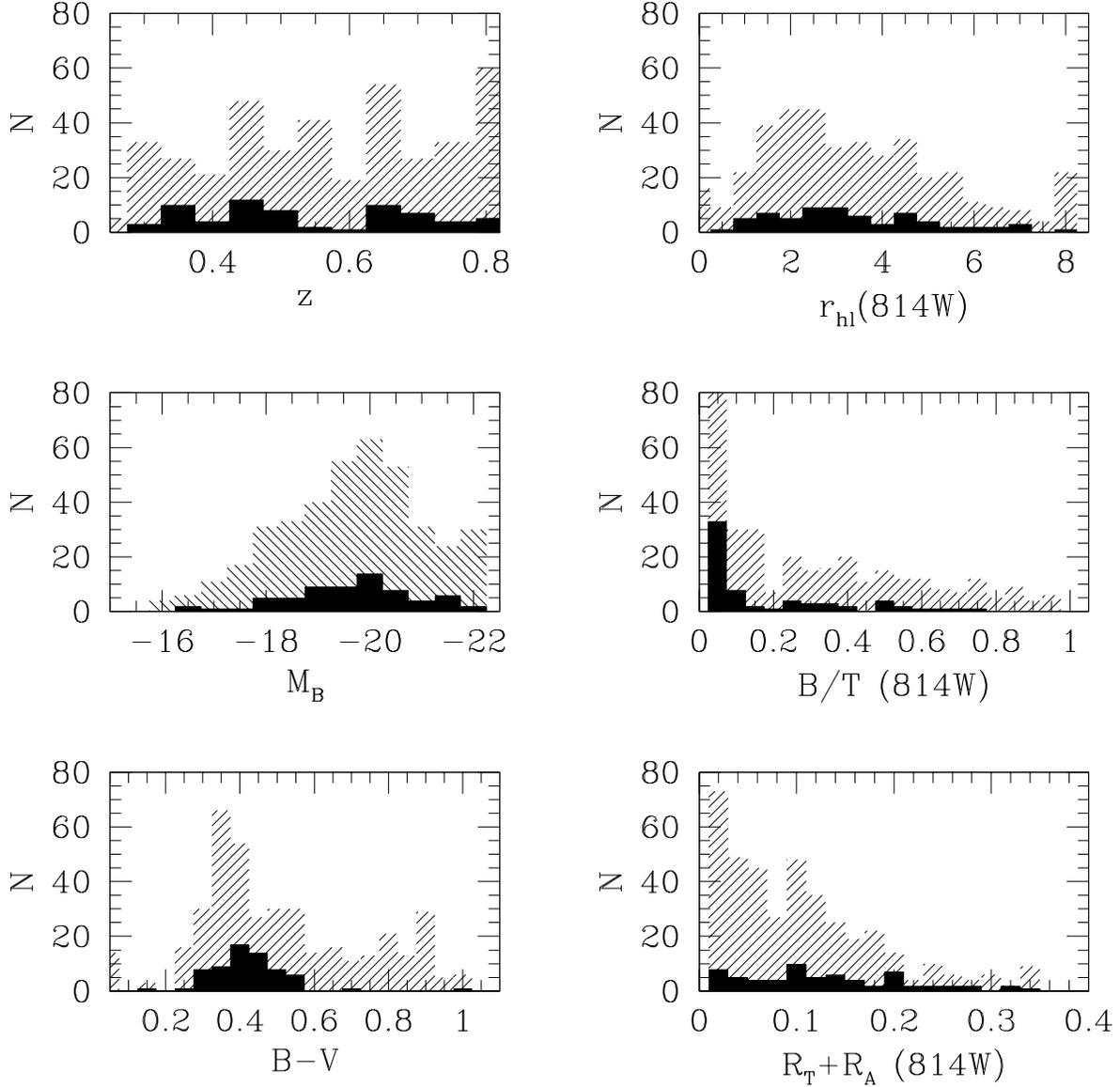}
\figcaption[hist1] {Histogram of 66 galaxies selected for chemical
analysis (filled) compared to the total 398 objects in the survey with
$0.26<z<0.82$.  We show the distribution as a function of redshift,
$M_B$, B-V color, half-light radius $R_{hl}$, bulge fraction
$F_{bulge}$, and asymmetry index, $R_T+R_A$. 
This figure demonstrates that galaxies selected as suitable for chemical analysis
are moderately representative of the larger DGSS sample in
the same redshift range
in terms of their redshift distributions,
sizes, luminosities and bulge fractions.  
However, the 66 selected galaxies preferentially 
have bluer B-V colors and more asymmetric 
morphologies, consistent with higher rates of star formation.  \label{hist} }
\end{figure}

\clearpage

\begin{figure} 
\plotone{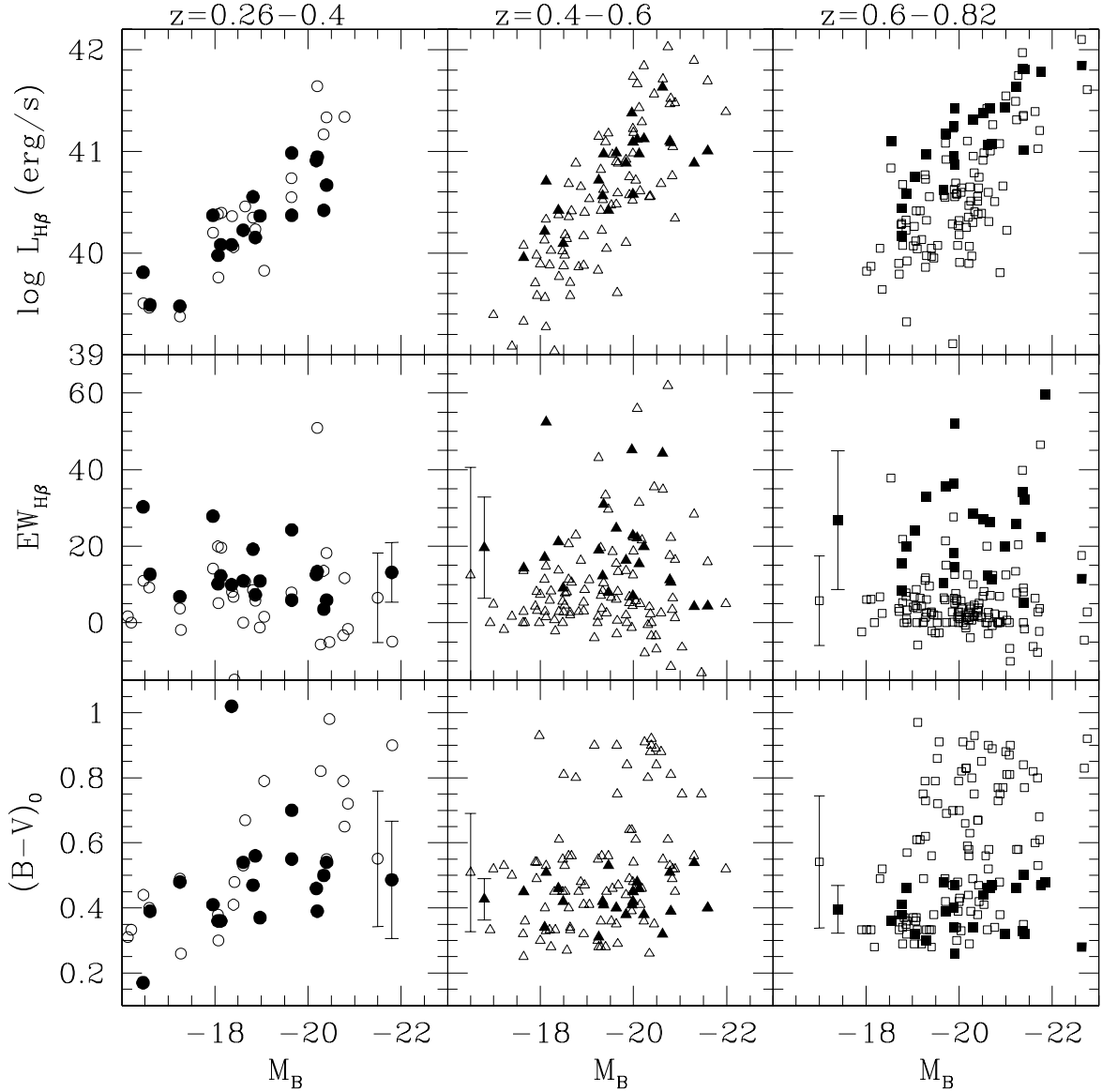}
\figcaption[select.ps] {A comparison of luminosity, color, and
emission line strength for 64 galaxies selected for chemical analysis
(solid symbols) to the 276 DGSS galaxies with
emission lines (open symbols) selected from among
the 398 DGSS galaxies in the three redshift bins spanning the
$0.26<z<0.82$ range.  Error bars show subsample means and dispersions.  
Galaxies selected for analysis in the lowest redshift bin are representative of
the DGSS objects in that redshift interval.  However, galaxies selected
for analysis from the highest redshift bins are biased toward
the bluest colors and highest emission line equivalent widths.
\label{select} } \end{figure}

\clearpage

\begin{figure}
\plotone{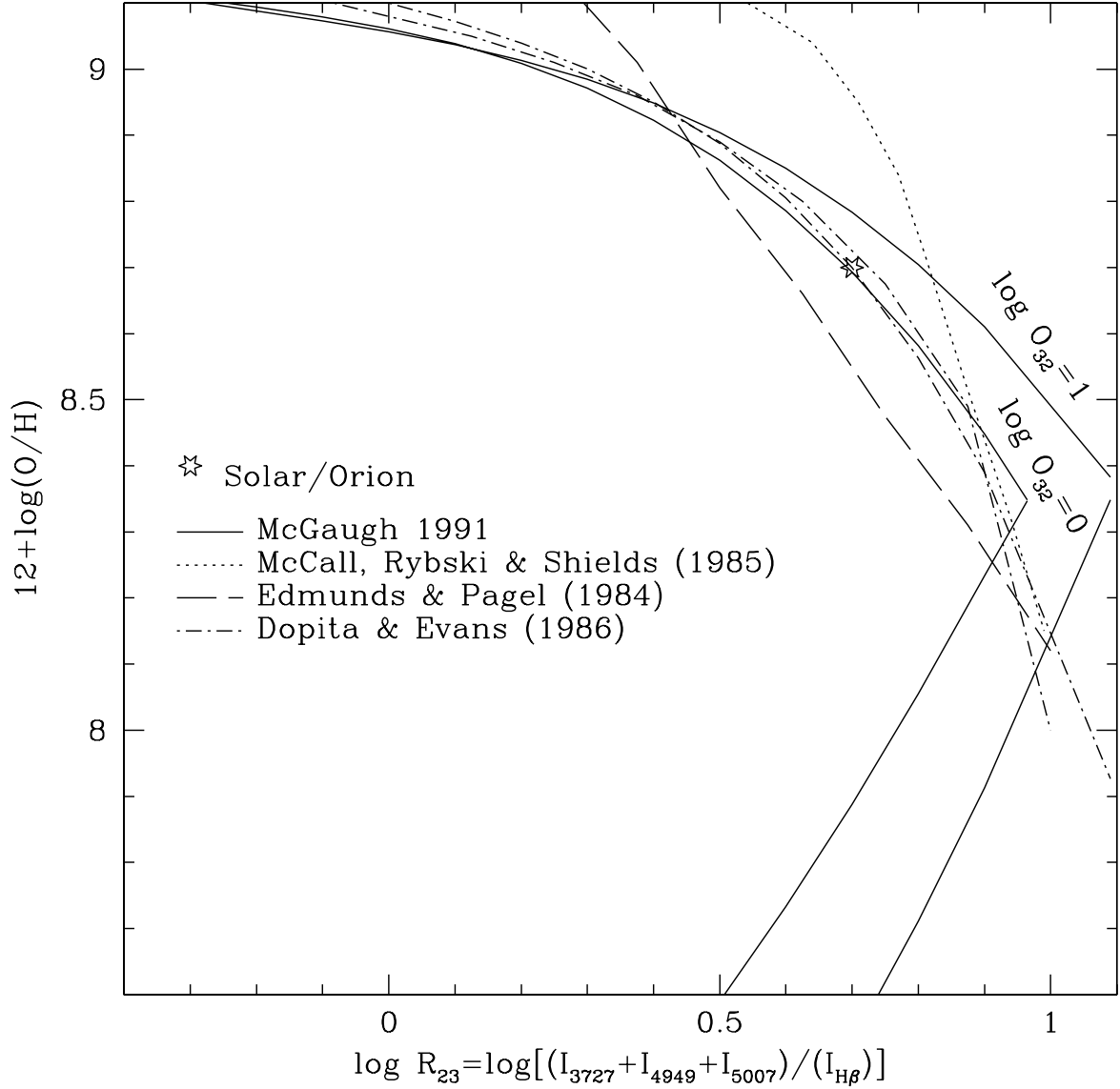}
\figcaption[R23OH.ps] {
Relation between the emission line ratio, $\log R_{23}\equiv
\log [ (I_{3727}+I_{4959}+I_{5007}) / (I_{H\beta}) ]$,
and oxygen abundance, 12+log(O/H) for
a variety of calibrations from the literature.
We adopt an analytical expression for the upper branch 
of the McGaugh (1991) relation which takes into account the ionization parameter
as measured by the $O_{32}$ ratio.  Two values of $\log(O_{32})=1$ and $\log(O_{32})=0$   
are shown. A star marks the Orion Nebula oxygen abundance  
(Walter, Dufour, \& Hester 1992) which is within 0.02 dex of
the recent solar abundance measurement by Prieto, Lambert, \& Asplund (2001).  
\label{R23OH} }
\end{figure}

\clearpage

\begin{figure}
\plotone{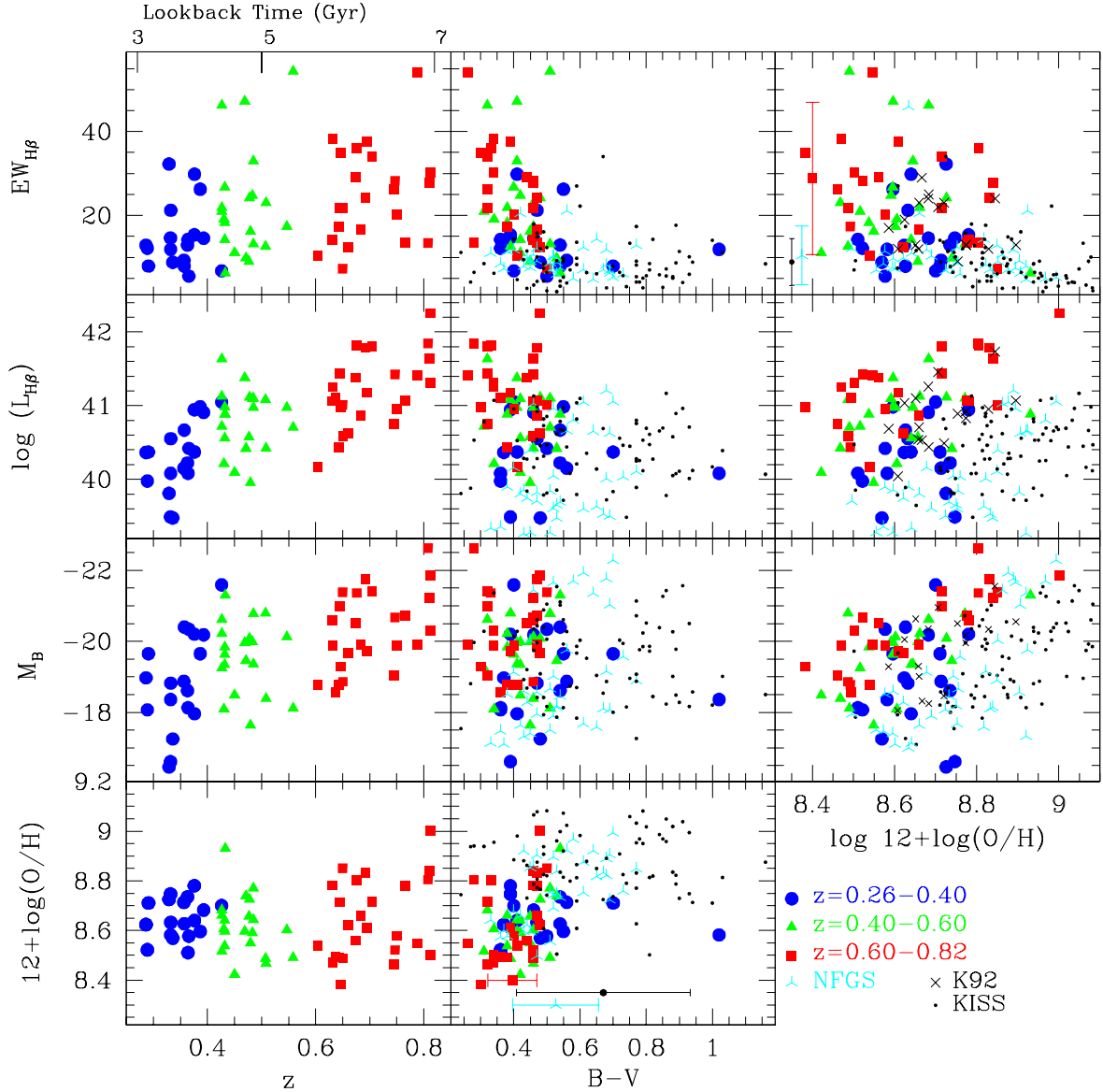}
\figcaption[zMBOH.ps]{Relation between redshift, $M_B$, 
$B-V$ color, $EW_{H\beta}$, $L_{H\beta}$, and 12+log(O/H)
for 64 DGSS galaxies with subsets of the local NFGS, K92, and KISS
samples chosen to match the DGSS emission-line ratio selection criteria.  
Points with error bars in some panels compare the means and dispersions
in color and $EW_{H\beta}$ for the DGSS, NFGS, and KISS samples.
DGSS galaxies in the highest redshift bin are preferentially
bluer with higher star formation rates and larger emission line equivalent
widths.  Each redshift interval exhibits
a correlation between metallicity and both blue and $H\beta$
luminosities, but with different zero points.  There is no 
significant correlation
between metallicity and color, or between metallicity and emission line
equivalent width. \label{zMBOH} }
\end{figure}

\clearpage

\begin{figure}
\plotone{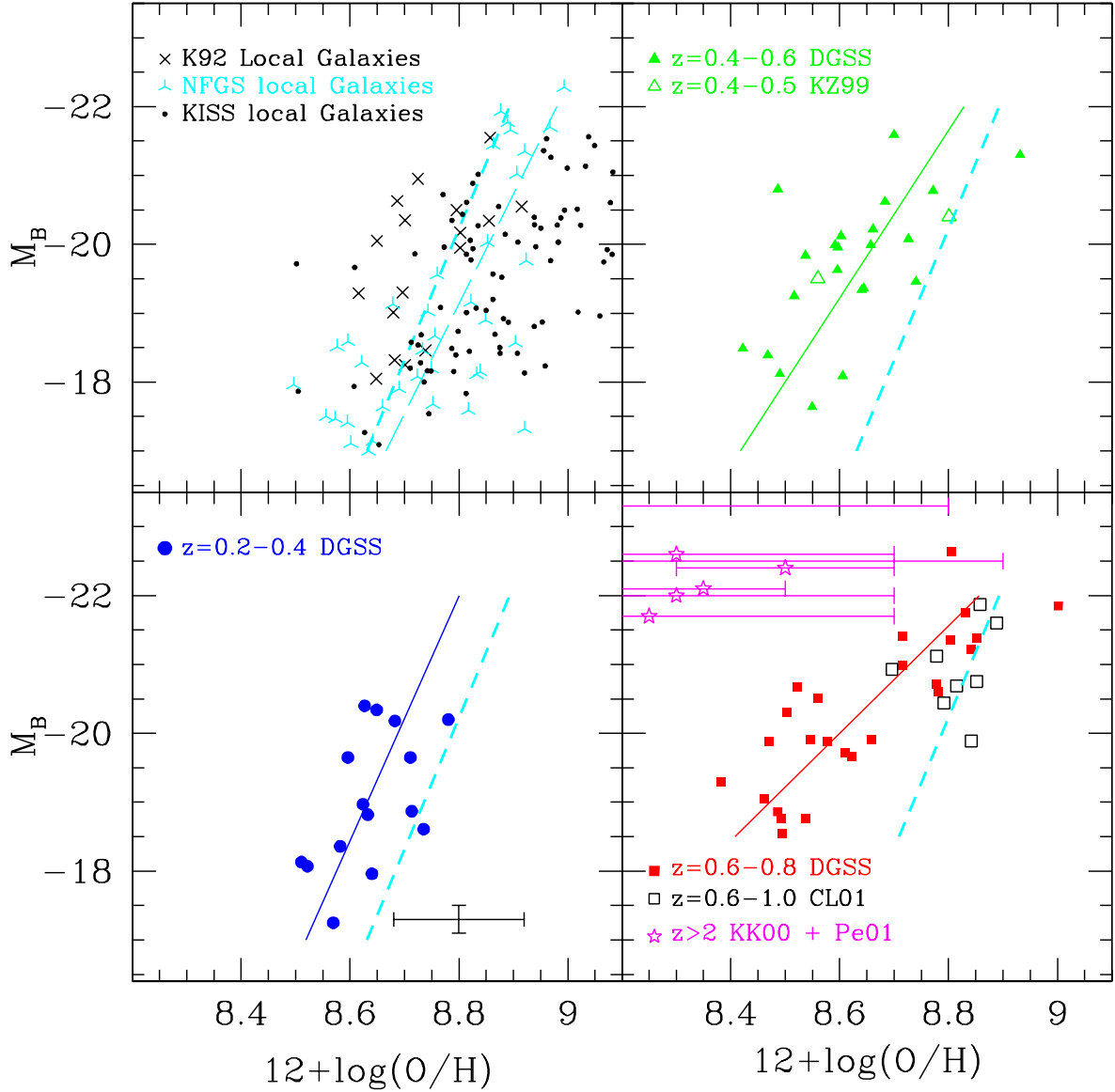}
\figcaption[LZ.ps] {Luminosity-metallicity (L-Z) relation
as a function of redshift including
samples of local $z<0.1$ galaxies from the NFGS, KISS, and Kennicutt
(1992b) catalogs, with intermediate-redshift DGSS objects from this
study (filled symbols) and the Carollo \& Lilly (2001) and Kobulnicky
\& Zaritsky (1999) compilations (open symbols).   A representative
error bar illustrating DGSS uncertainties on DGSS data 
is shown.  Stars denote high-redshift objects (Kobulnicky \&
Koo 2000; Pettini \etal\ 2001) with error bars representing the
full permitted range of metallicities. 
Lines show the best-fit relations including errors on both axes. 
The heavy dashed line in each panel is the
fit to local K92 and NFGS galaxies.  The long-dashed line
in the upper left is the fit to all K92, NFGS, and KISS galaxies.  
Table~3 lists the details of
the fit for each subsample.
The L-Z relation based on the DGSS galaxies is increasingly offset
from the local one with increasing redshift.  \label{LZ} } \end{figure}

\clearpage

\begin{figure} 
\plotone{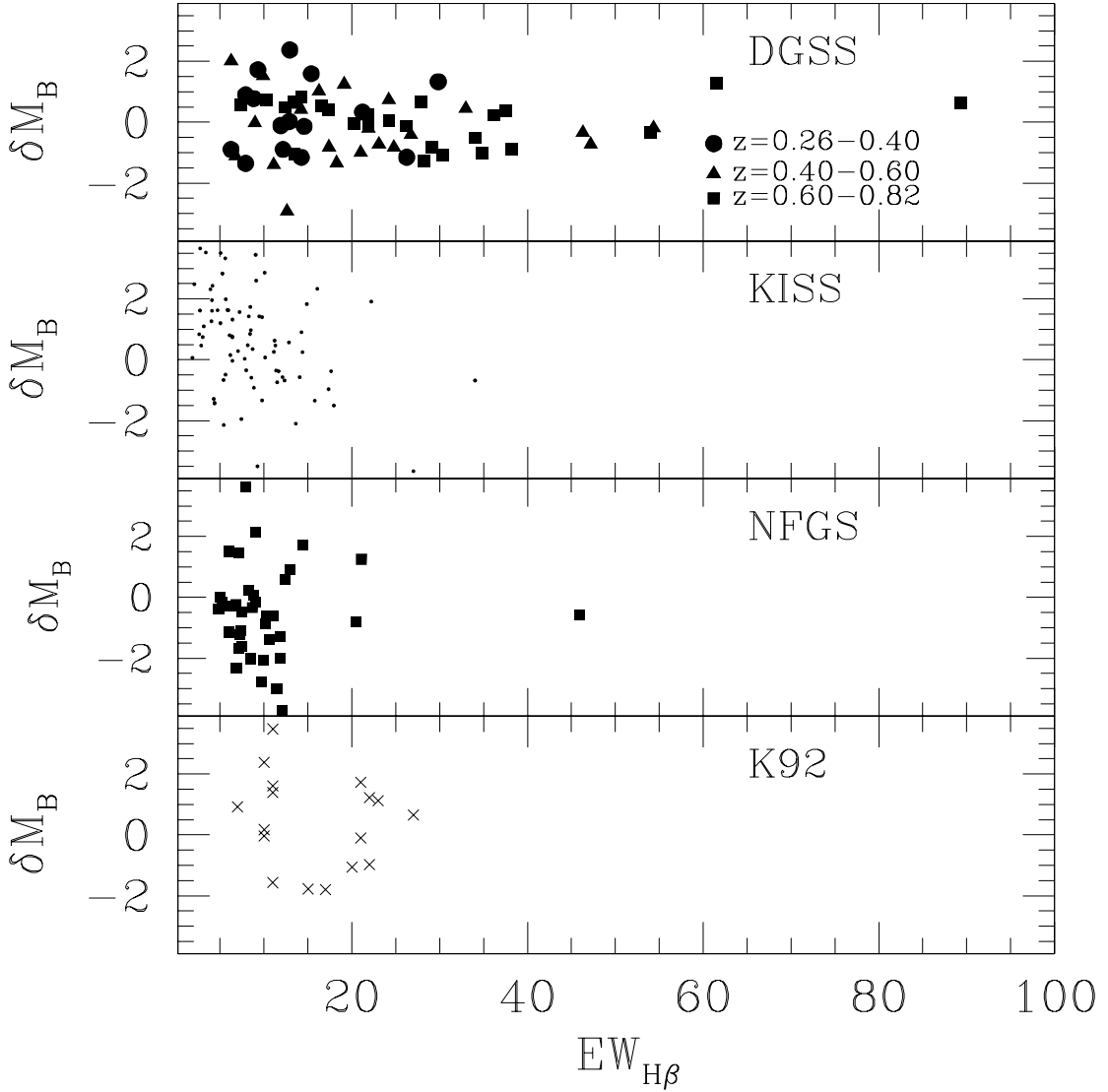}
\figcaption[LZresidEWHB.ps] {Galaxy $EW_{H\beta}$ versus luminosity residuals,
$\delta{M_B}$, from best-fit linear relations in the L-Z plane.
DGSS subsampels are fitted individually using the relations
of Table~2, while the the fit to the combined 
NFGS+KISS+K92 objects is used for those samples.  Note that KISS 
residuals are systematically
positive (faint) compared to the NFGS residuals.
There is no correlation between $EW_{H\beta}$
and L-Z residuals for the local samples.   There is only a weak
correlation among the DGSS galaxies in the highest redshift bin, 
driven mostly by the few galaxies
with extreme values of $EW_{H\beta}$. We searched for
other parameters, including galaxy color and size, which might correlate with
L-Z residuals and explain some of the dispersion in the
L-Z relation, but no significant correlations were found.
Dispersion in the L-Z relation is either intrinsic or
attributable to a yet unidentified combination of parameters.
\label{LZresidew} } \end{figure}

\clearpage

\begin{figure}
\plotone{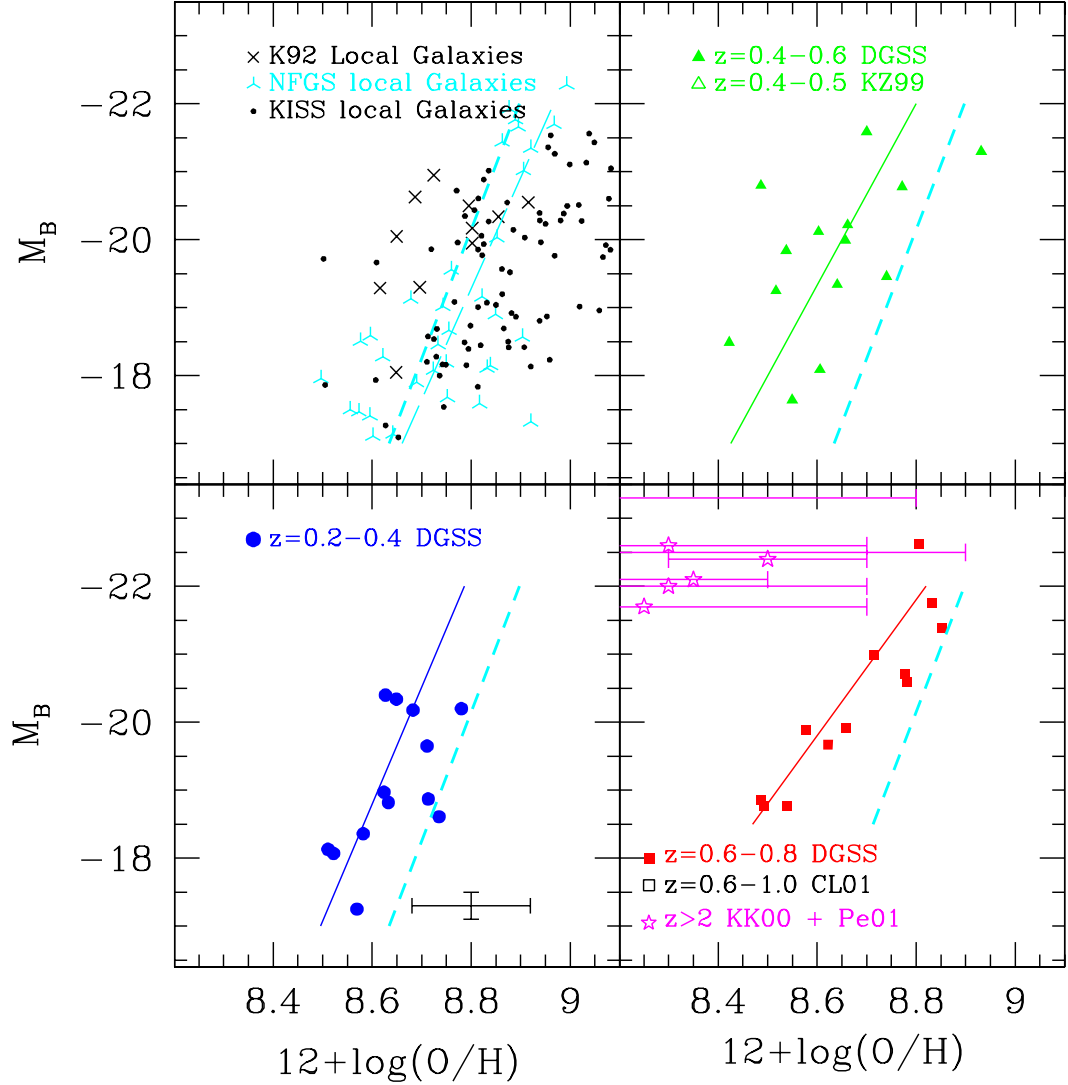}
\figcaption[LZlimEW.ps] {Luminosity-metallicity (L-Z) relations 
showing only local and DGSS galaxies with $EW_{H\beta}<20$ \AA.  
The mean L-Z relations 
for the local and distant samples fitted here are essentially unchanged
compared to the full set of galaxies in Figure~\ref{LZ},
indicating that emission line EW
differences between the local and DGSS samples are not responsible
for the observed offsets in the luminosity-metallicity relation. 
Here again,  DGSS galaxies are $\sim1-2$ mag
brighter than local galaxies of comparable metallicity.
 \label{LZlim} } \end{figure}

\clearpage

\begin{figure}
\plotone{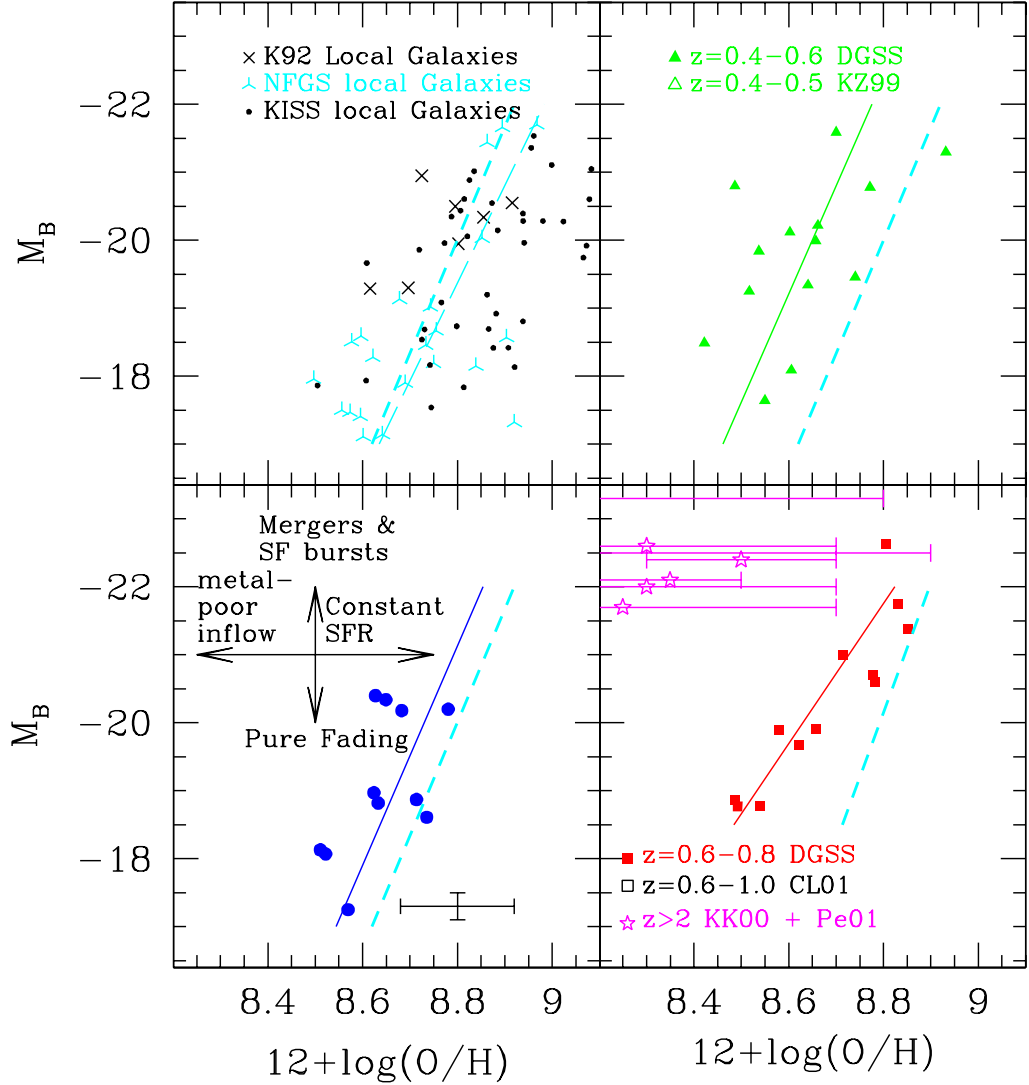}
\figcaption[LZlimBOTH.ps] {Luminosity-metallicity (L-Z) relation 
showing only local and DGSS galaxies with
 $B-V<0.6$ and $EW_{H\beta}<20$ \AA.  The best fit L-Z relations here are 
nearly unchanged
from the previous plots, indicating that color and emission line EW
differences between the local and DGSS samples are not responsible
for the offsets in the luminosity-metallicity relation.   
The schematic in the lower left panel indicates the evolution in the L-Z plane
caused by constant star formation, passive evolution, 
metal-poor gas inflow, and star formation bursts and/or 
galaxy mergers.  If DGSS galaxies evolve into
disklike star-forming galaxies today, then 
some combination of these processes are
responsible for evolving the $z=0.6-0.8$ 
galaxies into the region occupied by today's
$z=0$ galaxies.  \label{LZlimBOTH} } \end{figure}

\clearpage

\begin{figure} 
\plotone{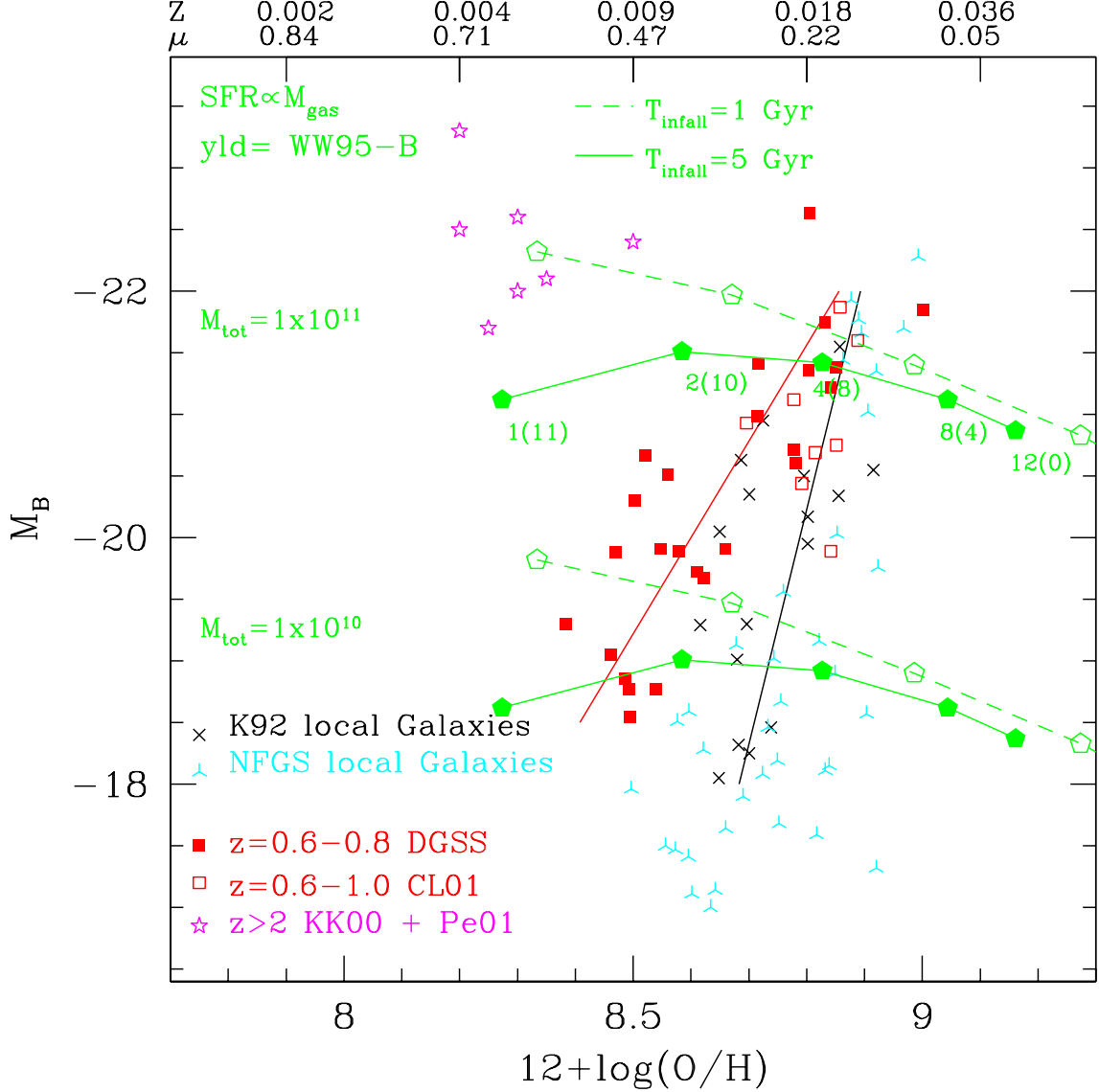}
\figcaption[LZevp2.cps] 
{L-Z relation with a subset of the symbols from
Figure~\ref{LZ}.  Lines show the $0.6<z<0.8$ (red) and local (black) L-Z relations
from Figure~\ref{LZ}.  Curved tracks and
 and pentagons show the chemical and luminous evolution of \textsc{P\'egase2} model galaxies
with a star formation rate proportional to the gas mass, where the
galaxy is built by exponentially-decreasing infall of primordial gas
with infall timescales of 1 Gyr and 5 Gyr.  Pentagons denote galaxies at
ages of 1, 2, 4, 8, and 12 Gyr 
(corresponding to lookback times, as shown in parentheses, of 11, 10, 8, 4, and 0 Gyr,
respectively). Galaxy masses of $10^{10}$ \mo\
and $10^{11}$ \mo\ (neglecting dark matter) are shown.  The top of the figure shows the
corresponding gas metallicity, $Z$, and mass fraction, $\mu$, assuming 
a closed-box scenario for an effective yield of 0.016 from Weaver \& 
Woosley (1995) series B models. 
These  models match the observed rapid rise in gas-phase metallicity of Lyman break galaxies 
but overproduce metals today compared to observed galaxies.
\label{LZevp2} } \end{figure}
\clearpage

\begin{figure} 
\plotone{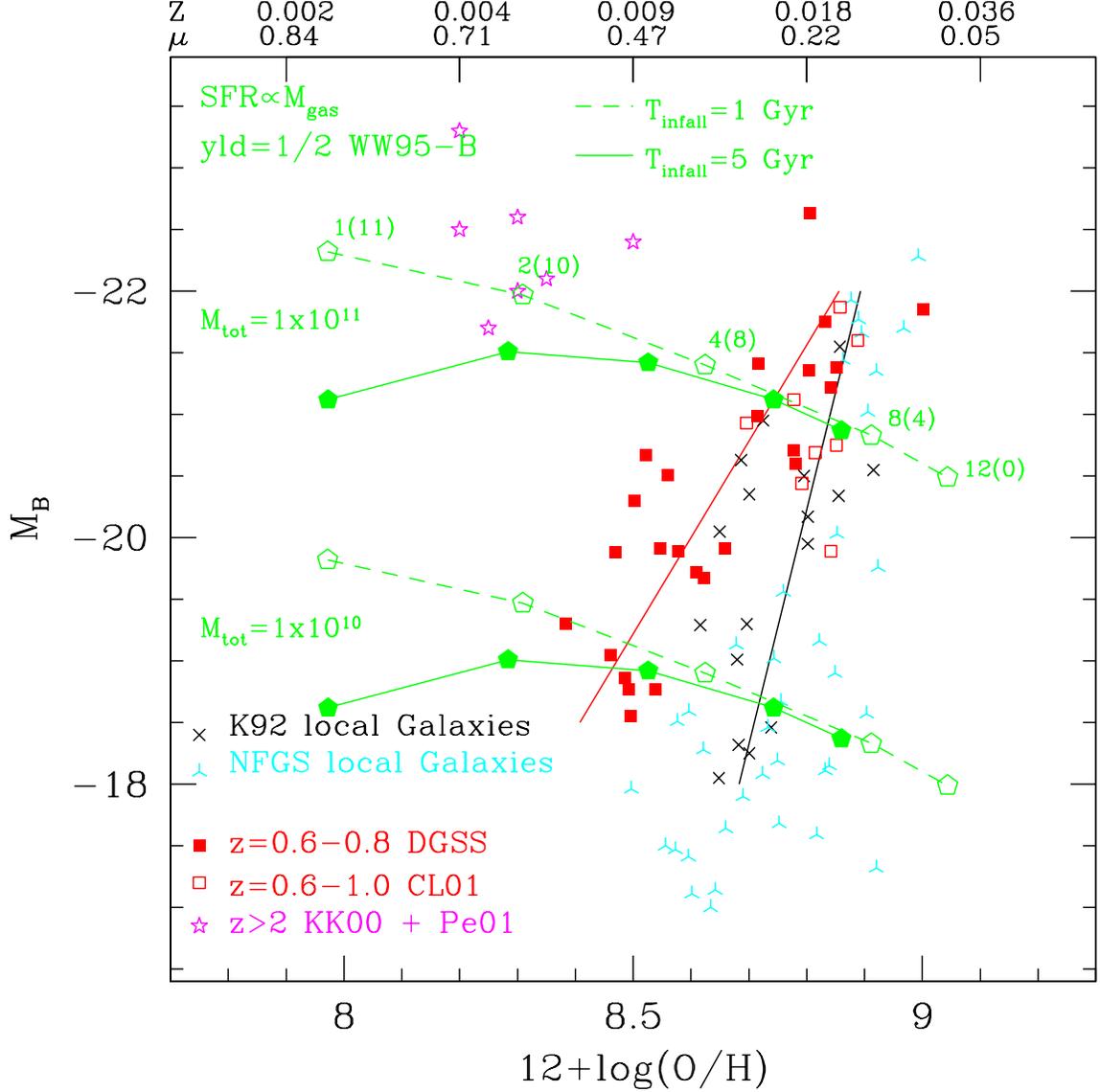}
\figcaption[LZevp.cps] 
{Comparison of data to evolutionary models in the L-Z plane similar to
Figure~\ref{LZevp2}, except that the effective yield of the models
has been arbitrarily reduced by a factor of 2 ($\eta_Z=0.5$).  
 Lines show the best-fit L-Z relations
from Figure~\ref{LZ}.  Curved tracks and
 and pentagons show the evolution of a model galaxy
with a star formation rate proportional to the gas mass where the
galaxy is built by exponentially-decreasing infall of primordial gas
with infall timescale of 1 Gyr and 5 Gyr, same as for the models in Figure~\ref{LZevp2}.   
Pentagons denote galaxies at
ages of 1, 2, 4, 8, and 12 Gyr 
(corresponding to lookback times, as shown in parentheses, of 11, 10, 8, 4, and 0 Gyr,
respectively).  Galaxy masses of $10^{10}$ \mo\
and $10^{11}$ \mo\ are shown.
The metallicities of the reduced-yield models do not
rise as quickly, do not overproduce metals at
late times and are in overall better agreement with the data.  
Longer gas infall timescales for low-mass galaxies compared to high-mass galaxies
gas is one way of reproducing the slope and offset variations of the 
local and distant L-Z relations.  Later formation epochs for low-mass galaxies
or reduced effective metal yields for low-mass galaxies
will also serve to produce the change in slope and offset for the
low-luminosity DGSS galaxies.  
\label{LZevp} } \end{figure}

\clearpage

\begin{figure} 
\plotone{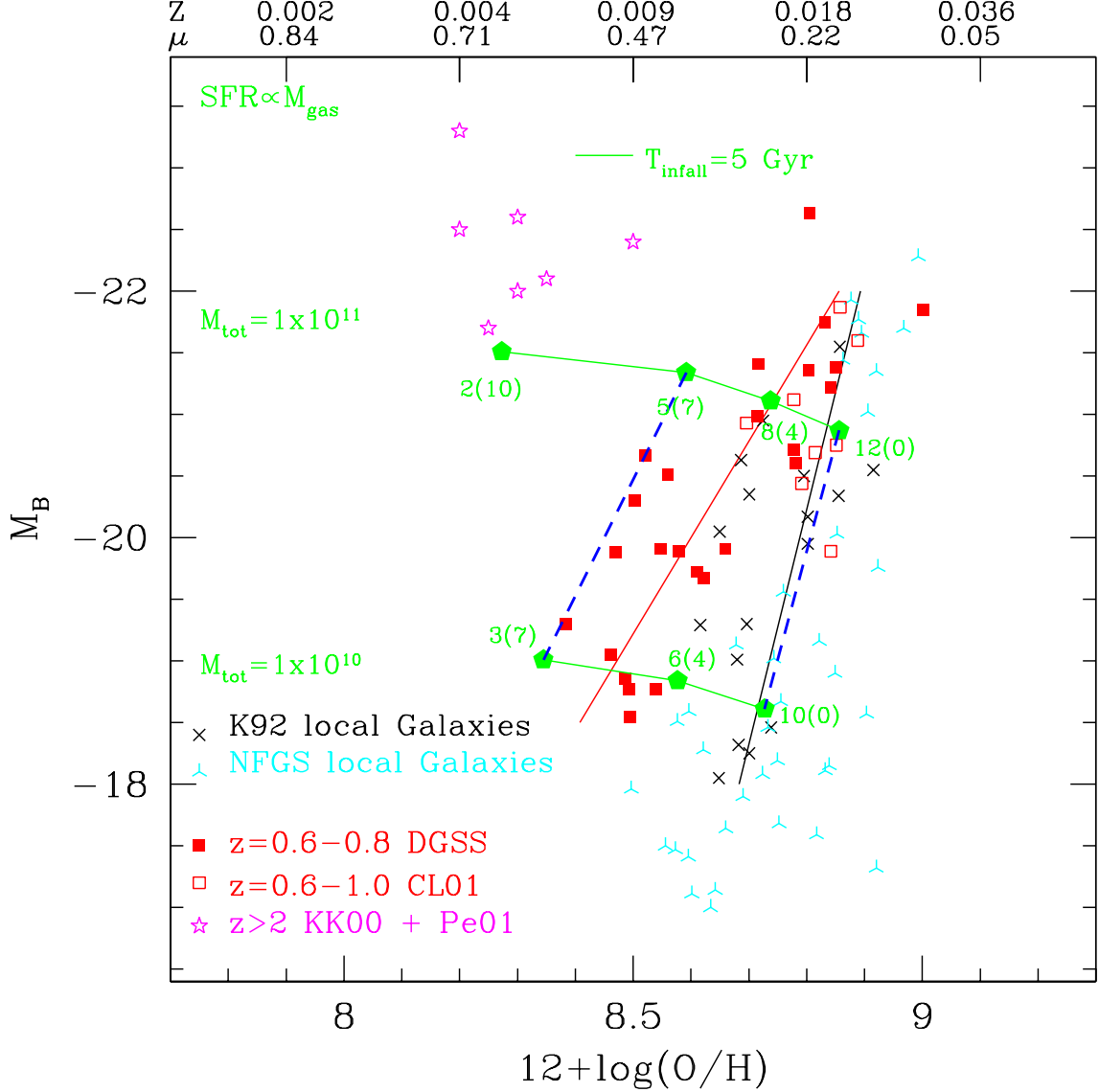}
\figcaption[LZevp4.cps] 
{Comparison of data to evolutionary models in the L-Z plane similar to
Figure~\ref{LZevp}, except that the effective yield of the $10^{10}~M_\odot$
model has been arbitrarily reduced by an additional factor of 20\% relative to the 
$10^{11}~M_\odot$ model to simulate the impact of increasingly efficient 
galactic winds in lower mass galaxies.  
 Lines show the best-fit L-Z relations
from Figure~\ref{LZ}. 
Here, the model points are labeled with age (and lookback times in parentheses) 
assuming that the
$10^{11}~M_\odot$ galaxy begins assembly at $T_0=1$ Gyr and that 
the $10^{11}~M_\odot$ galaxy begins assembly 
at $T_0=3$ Gyr.  Heavy dotten lines (blue) connect model points at the same lookback time.
The combined effects of decreased yield and
later formation epoch for the lower mass galaxy qualitatively
simulates both the slope of the L-Z relation {\it and} the
change in slope and offset of the L-Z relation at earlier epochs.
However, the models slightly underpredict the chemical
enrichment observed in galaxies at $0.6<z<0.8$, and they do a poor job of predicting
the evolution of luminous galaxies.   
\label{LZevp4} } \end{figure}

\clearpage

\begin{figure} 
\plotone{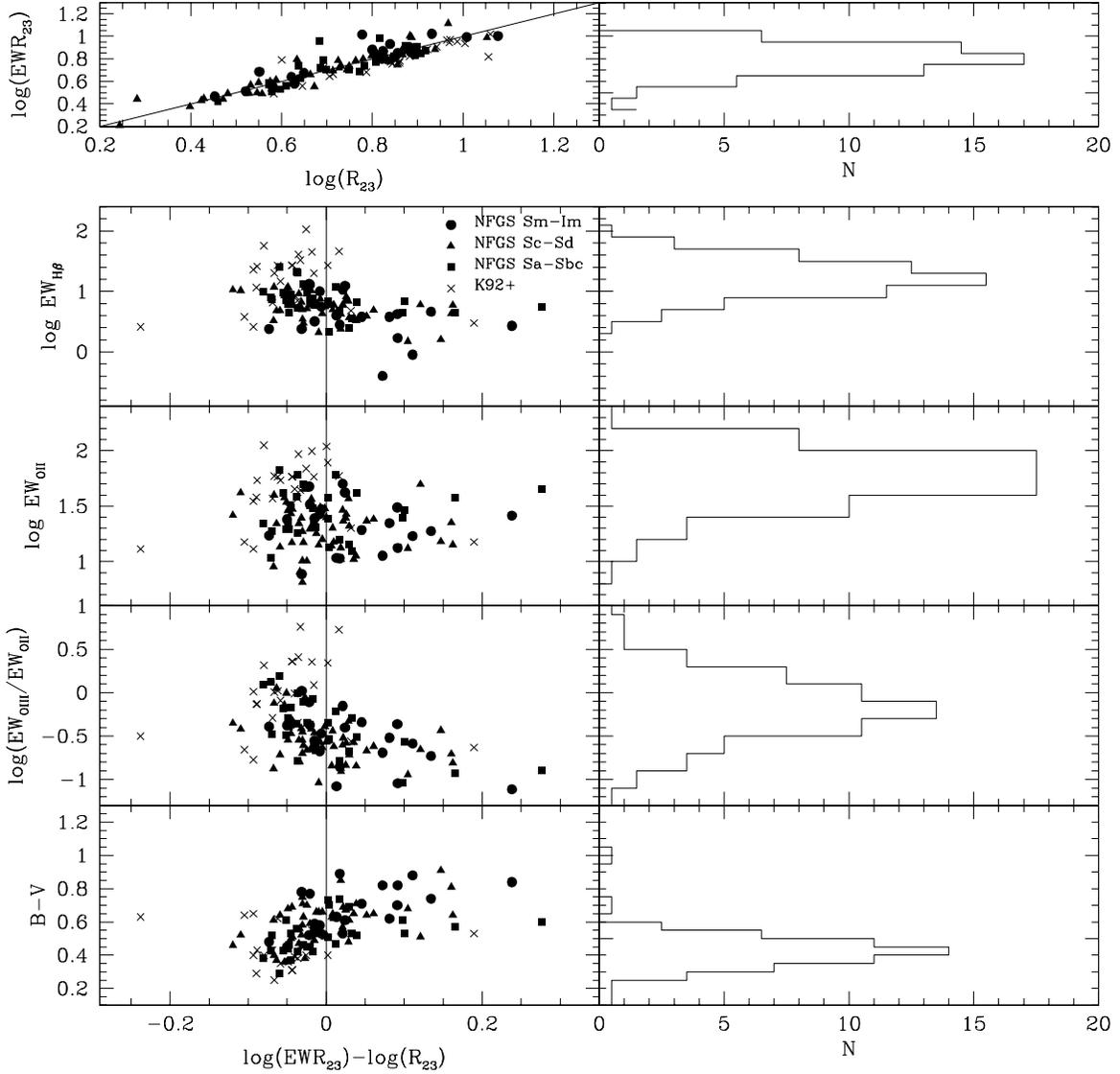}
\figcaption[histcompare.ps] 
{Comparison of the quantity $R_{23}$ with $EWR_{23}$ from
Kobulnicky \& Phillips (2003).  The strong
correlation between $R_{23}$ and $EWR_{23}$ (upper left panel)
shows that 
oxygen abundances can be estimated from equivalent width ratios at
least as well as from dereddened line fluxes. 
The RMS dispersion from the 1-to-1 relation is
$\sigma(log[R_{23}]) = 0.07$ dex.   Lower panels explore
the residuals in the correlation as a function of line strength, line
ratio, and galaxy color.  Histograms in the right column
show the distribution of colors and emission line strengths for
the 66 emission-line
galaxies selected for analysis from the DGSS.   The majority
of galaxies lie in regimes where the $EWR_{23}$ vs. $R_{23}$
correlation is strong and well-behaved.
\label{histcompare} } \end{figure}

\clearpage

\begin{deluxetable}{rrrrcccccccrrrrrrrl}
\label{srctable.tab}
\rotate
\tabletypesize{\scriptsize} 
\setlength{\tabcolsep}{0.02in} 
\tablewidth{8.3in}
\tablecaption{DGSS Selected Galaxies}
\tablehead{
\colhead{ID } & 
\colhead{Object} & 
\colhead{RA (2000)}   & 
\colhead{DEC (2000)}   & 
\colhead{z}   & 
\colhead{$I_{814} AB$} & 
\colhead{$M_B$} & 
\colhead{$(B-V)$} &
\colhead{$R_{hl}$} &
\colhead{$B/T$}  &
\colhead{$R_T$}   &
\colhead{$R_A$}   &
\colhead{$EW_{3727}$}   &
\colhead{$EW_{4861}$}   &
\colhead{$EW_{4959}$}     &
\colhead{$EW_{5007}$}  &
\colhead{$12+log(O/H)$}  &
\colhead{$SFR$}  &
\colhead{Notes}  \\ [.2ex]
\colhead{(1)} &  
\colhead{(2)} &  
\colhead{(3)} &  
\colhead{(4)} &  
\colhead{(5)} &
\colhead{(6)} &
\colhead{(7)} &
\colhead{(8)} &
\colhead{(9)} &
\colhead{(10)} &
\colhead{(11)} &
\colhead{(12)} &
\colhead{(13)} &
\colhead{(14)} &
\colhead{(15)} &
\colhead{(16)}  &
\colhead{(17)}  &
\colhead{(18)}  &
\colhead{(19)}   } 
\startdata
 1  &  073-2658 &  14:17:47.597 & 52:29:03.36 & 0.2863 &  21.13 & -18.9 &  0.34 &  2.88 &  0.28 &  0.04 &  0.10 &  62.1$\pm$1.6 &  10.8$\pm$1.1 &  11.1$\pm$2.2 &  23.0$\pm$1.2 & 8.62$\pm$0.11 &   0.7 & A,B \\
 2 &  072-2273 &  14:17:38.928 & 52:29:34.72 & 0.2888 &  22.03 & -18.0 &  0.36 &  2.36 &  0.00 &  0.03 &  0.05 &  49.5$\pm$1.4 &  10.1$\pm$0.4 &   6.5$\pm$0.4 &  25.7$\pm$0.4 & 8.52$\pm$0.08 &   0.3 & A \\
 3 &  072-4040 &  14:17:37.838 & 52:28:58.44 & 0.2908 &  20.01 & -19.6 &  0.70 &  9.42 &  0.00 &  0.15 &  0.06 &  26.9$\pm$1.1 &   5.9$\pm$0.3 &   1.7$\pm$0.2 &   5.7$\pm$0.2 & 8.71$\pm$0.08 &   1.0 & A \\
 4 &  172-1242 &  14:16:35.570 & 52:17:27.91 & 0.3288 &  24.22 & -16.4 &  0.17 &  1.37 &  0.10 &  -.02 &  0.01 &  22.9$\pm$1.7 &  30.2$\pm$3.6 &  39.1$\pm$3.9 &    121$\pm$4  & 8.73$\pm$0.04 &   0.1 & A,C \\
 5 &  092-1375 &  14:17:26.699 & 52:27:19.48 & 0.3315 &  23.84 & -16.6 &  0.39 &  1.54 &  0.15 &  0.06 &  0.04 &  22.8$\pm$1.2 &  12.6$\pm$0.7 &  13.2$\pm$0.7 &  32.5$\pm$0.7 & 8.75$\pm$0.03 &   0.1 & A,C \\
 6 &  134-2322 &  14:17:05.547 & 52:21:18.17 & 0.3319 &  21.28 & -18.3 &  1.02 &  2.39 &  0.34 &  0.05 &  0.15 &  39.1$\pm$1.0 &   9.9$\pm$0.5 &  10.8$\pm$0.6 &  24.1$\pm$0.5 & 8.58$\pm$0.06 &   0.7 & A \\
 7 &  142-4644 &  14:16:51.341 & 52:20:52.38 & 0.3322 &  21.52 & -18.8 &  0.47 &  3.62 &  0.00 &  0.03 &  0.01 &  76.1$\pm$1.2 &  19.2$\pm$1.6 &   6.9$\pm$1.9 &  30.5$\pm$2.1 & 8.63$\pm$0.09 &   1.2 & A \\
 8 &  223-2132 &  14:16:07.432 & 52:11:25.04 & 0.3359 &  23.16 & -17.2 &  0.48 &  1.16 &  0.04 &  -.01 &  0.00 &  33.8$\pm$2.8 &   6.8$\pm$0.9 &   5.1$\pm$0.8 &  15.1$\pm$1.1 & 8.57$\pm$0.09 &   0.1 & A \\
 9 &  124-3864 &  14:17:14.790 & 52:21:50.38 & 0.3564 &  21.55 & -18.8 &  0.56 &  3.02 &  0.00 &  0.05 &  0.07 &  34.2$\pm$3.3 &   7.2$\pm$0.5 &   0.8$\pm$0.7 &   2.5$\pm$0.5 & 8.71$\pm$0.10 &   0.5 & A \\
  10 &  173-5210 &  14:16:36.664 & 52:17:39.89 & 0.3570 &  20.03 & -20.4 &  0.54 &  4.51 &  0.05 &  0.07 &  0.06 &  34.9$\pm$0.9 &   5.8$\pm$0.4 &   1.4$\pm$0.4 &   3.5$\pm$0.4 & 8.62$\pm$0.12 &   1.7 & A \\
11 &  103-6061 &  14:17:27.434 & 52:26:08.93 & 0.3631 &  21.88 & -18.6 &  0.51 &  0.93 &  0.49 &  0.03 &  0.01 &  38.9$\pm$0.7 &  10.9$\pm$0.6 &   4.4$\pm$0.5 &  11.3$\pm$0.5 & 8.74$\pm$0.07 &   0.6 & A \\
12 &  203-4851 &  14:16:21.680 & 52:14:14.78 & 0.3639 &  22.58 & -18.1 &  0.36 &  1.18 &  0.70 &  0.02 &  0.01 &  52.6$\pm$1.1 &  12.2$\pm$0.4 &  13.8$\pm$0.4 &  34.6$\pm$0.4 & 8.51$\pm$0.06 &   0.3 & A \\
  13 &  213-6640 &  14:16:13.664 & 52:13:20.44 & 0.3656 &  20.22 & -20.3 &  0.50 &  4.62 &  0.03 &  0.06 &  0.05 &  26.2$\pm$3.2 &   4.2$\pm$0.3 &   0.5$\pm$0.3 &   0.8$\pm$0.3 & 8.66$\pm$0.15 &   0.9 & A,B \\
14 &  313-3557 &  14:15:11.917 & 52:01:10.88 & 0.3757 &  20.56 & -20.2 &  0.42 &  2.35 &  0.25 &  0.06 &  0.04 &  44.4$\pm$0.9 &  13.3$\pm$0.2 &   3.3$\pm$0.1 &   9.0$\pm$0.2 & 8.78$\pm$0.07 &   2.8 & A \\
15 &  304-4028 &  14:15:17.787 & 52:01:23.63 & 0.3760 &  22.78 & -17.9 &  0.41 &  1.38 &  0.02 &  0.03 &  0.04 &  99.5$\pm$1.1 &  27.8$\pm$1.3 &  14.6$\pm$1.6 &    46$\pm$1.6 & 8.64$\pm$0.07 &   0.7 & A \\
  16 &  183-6415 &  14:16:30.453 & 52:16:42.93 & 0.3865 &  21.00 & -19.6 &  0.55 &  7.06 &  0.00 &  0.14 &  0.07 &     109$\pm$50. &  24.2$\pm$2.5 &   8.3$\pm$2.5 &  27.8$\pm$2.7 & 8.60$\pm$0.12 &   3.6 & A,B \\
17  &  193-5319 &  14:16:24.702 & 52:15:23.83 & 0.3929 &  20.62 & -20.1 &  0.46 &  4.45 &  0.03 &  0.06 &  0.07 &  55.5$\pm$0.4 &  12.5$\pm$0.2 &   2.3$\pm$0.2 &   7.3$\pm$0.4 & 8.69$\pm$0.09 &   2.8 & A \\
  18 &  092-1962 &  14:17:26.421 & 52:27:06.04 & 0.4261 &  19.49 & -21.5 &  0.40 &  6.35 &  0.03 &  0.20 &  0.15 &  24.4$\pm$0.2 &   4.8$\pm$0.2 &   1.0$\pm$0.1 &   4.3$\pm$0.1 & 8.68$\pm$0.09 &   3.3 & A \\
19 &  304-3354 &  14:15:17.610 & 52:00:56.86 & 0.4263 &  20.57 & -20.6 &  0.29 &  1.90 &  0.51 &  0.09 &  0.12 &     113$\pm$2 &  44.3$\pm$2.4 &  30.0$\pm$2.6 &  97.1$\pm$2.6 & 8.68$\pm$0.05 &  12.7 & A \\
20 &  223-6341 &  14:16:07.359 & 52:12:07.83 & 0.4263 &  21.94 & -19.2 &  0.29 &  2.71 &  0.06 &  0.06 &  0.07 &  82.1$\pm$2.2 &    19$\pm$2.2 &  14.4$\pm$1.1 &  48.1$\pm$1.1 & 8.52$\pm$0.09 &   1.5 & A \\
21 &  292-3870 &  14:15:14.882 & 52:03:47.19 & 0.4266 &  20.90 & -20.2 &  0.36 &  2.98 &  0.00 &  0.19 &  0.08 &  63.6$\pm$2.2 &  19.9$\pm$2.2 &  11.9$\pm$1.1 &  40.0$\pm$1.1 & 8.66$\pm$0.07 &   4.2 & A \\
22 &  093-6667 &  14:17:34.454 & 52:27:25.05 & 0.4306 &  23.08 & -18.0 &  0.34 &  3.07 &  0.01 &  -.00 &  0.01 &  39.5$\pm$0.9 &  17.1$\pm$0.5 &  28.2$\pm$0.5 &  57.1$\pm$0.5 & 8.61$\pm$0.03 &   0.5 & A \\
23 &  164-3515 &  14:16:47.287 & 52:17:57.62 & 0.4320 &  21.74 & -19.3 &  0.42 &  3.23 &  0.01 &  0.02 &  0.02 &  54.8$\pm$0.8 &  12.2$\pm$1.1 &   3.2$\pm$1.3 &  14.3$\pm$1.3 & 8.64$\pm$0.05 &   1.2 & A \\
24 &  164-2417 &  14:16:46.114 & 52:17:53.61 & 0.4323 &  21.30 & -19.8 &  0.38 &  3.86 &  0.08 &  0.08 &  0.08 &  78.4$\pm$0.8 &  16.2$\pm$1.7 &   8.5$\pm$2.0 &  29.2$\pm$2.0 & 8.53$\pm$0.10 &   2.4 & A \\
25 &  074-4757 &  14:17:48.165 & 52:27:48.29 & 0.4323 &  21.49 & -19.6 &  0.40 &  4.82 &  0.03 &  0.15 &  0.09 &     103$\pm$2 &  24.7$\pm$1.5 &  11.6$\pm$1.3 &  38.8$\pm$1.5 & 8.60$\pm$0.08 &   3.1 & A \\
26 &  094-1054 &  14:17:31.025 & 52:25:24.32 & 0.4331 &  19.66 & -21.3 &  0.54 &  5.37 &  0.10 &  0.09 &  0.11 &  11.5$\pm$0.6 &   4.2$\pm$0.2 &   0.6$\pm$0.4 &   2.0$\pm$0.2 & 8.93$\pm$0.01 &   2.8 & A \\
  27 &  313-7545 &  14:15:09.713 & 52:01:46.44 & 0.4503 &  22.70 & -18.4 &  0.42 &  2.54 &  0.01 &  0.01 &  0.01 & 54.9$\pm$12.1 &   9.1$\pm$0.8 &   6.8$\pm$1.1 &  24.6$\pm$1.1 & 8.42$\pm$0.10 &   0.4 & A,B \\
28 &  183-4770 &  14:16:36.632 & 52:16:38.91 & 0.4691 &  21.38 & -19.9 &  0.41 &  1.85 &  0.56 &  0.05 &  0.06 &     121$\pm$1 &  45.1$\pm$3.0 &  53.7$\pm$3.2 & 126$\pm$3     & 8.60$\pm$0.05 &   7.7 & A \\
29 &  223-7714 &  14:16:04.136 & 52:12:15.01 & 0.4711 &  21.72 & -19.4 &  0.53 &  2.12 &  0.31 &  0.02 &  0.04 &    32$\pm$1.3 &   7.8$\pm$0.2 &     2$\pm$0.2 &   6.0$\pm$0.3 & 8.73$\pm$0.07 &   0.9 & A \\
30 &  062-1570 &  14:17:46.321 & 52:30:43.30 & 0.4770 &  21.38 & -19.9 &  0.45 &  2.49 &  0.34 &  0.05 &  0.06 &  26.9$\pm$0.6 &   6.9$\pm$0.4 &   4.6$\pm$0.4 &  15.9$\pm$0.4 & 8.65$\pm$0.05 &   1.2 & A \\
31 &  162-5547 &  14:16:37.359 & 52:18:33.42 & 0.4776 &  21.24 & -20.0 &  0.48 &  2.03 &  0.03 &  0.06 &  0.05 &  73.2$\pm$0.8 &  22.1$\pm$0.6 &   7.3$\pm$0.6 &  24.5$\pm$0.6 & 8.73$\pm$0.06 &   4.5 & A \\
32 &  083-6536 &  14:17:37.724 & 52:28:28.13 & 0.4798 &  23.75 & -17.6 &  0.45 &  0.44 &  0.26 &  0.03 &  -.02 &  55.5$\pm$1.7 &  14.2$\pm$1.2 &  13.6$\pm$1.3 &  38.9$\pm$1.7 & 8.55$\pm$0.07 &   0.2 & A \\
33 &  212-2260 &  14:16:08.191 & 52:13:03.10 & 0.4832 &  21.43 & -19.9 &  0.42 &  2.77 &  0.14 &  0.08 &  0.08 &  94.1$\pm$0.6 &  22.7$\pm$0.2 &  11.7$\pm$0.2 &  38.9$\pm$0.2 & 8.59$\pm$0.08 &   3.9 & A \\
34 &  303-1256 &  14:15:18.750 & 52:01:58.57 & 0.4850 &  22.09 & -19.3 &  0.41 &  5.05 &  0.52 &  0.05 &  0.08 &  98.5$\pm$4.3 &  30.9$\pm$2.5 &  16.8$\pm$2.8 &  65.3$\pm$2.8 & 8.64$\pm$0.07 &   3.0 & A \\
35 &  303-3546 &  14:15:17.146 & 52:02:18.14 & 0.4850 &  20.52 & -20.7 &  0.51 &  4.96 &  0.10 &  0.04 &  0.06 &  41.6$\pm$0.4 &  11.2$\pm$0.4 &   1.6$\pm$0.5 &   5.1$\pm$0.5 & 8.77$\pm$0.08 &   4.6 & A \\
  36 &  183-1153 &  14:16:35.713 & 52:16:00.42 & 0.5070 &  20.74 & -20.8 &  0.39 &  5.96 &  0.42 &  0.09 &  0.13 &  64.3$\pm$20.0 &  10.6$\pm$0.6 &   3.6$\pm$0.6 &  15.5$\pm$0.6 & 8.49$\pm$0.12 &   3.9 & A,B \\
37 &  262-5149 &  14:15:33.122 & 52:06:56.39 & 0.5084 &  23.14 & -18.3 &  0.46 &  2.64 &  0.08 &  0.02 &  -.03 &  70.6$\pm$1.0 &  21.0$\pm$2.3 &  25.5$\pm$1.9 &    92$\pm$1.9 & 8.47$\pm$0.06 &   0.8 & A \\
38 &  114-3114 &  14:17:19.113 & 52:23:47.26 & 0.5470 &  21.59 & -20.1 &  0.46 &  3.43 &  0.03 &  0.03 &  0.04 &  78.8$\pm$2.0 &  15.3$\pm$1.0 &   3.1$\pm$1.0 &   6.1$\pm$1.0 & 8.60$\pm$0.12 &   3.1 & A \\
39 &  283-3961 &  14:15:31.425 & 52:04:46.65 & 0.5586 &  23.57 & -18.1 &  0.51 &  1.07 &  0.63 &  0.04 &  0.03 &   134$\pm$2.2 &  52.3$\pm$8.6 & 71.2$\pm$10.4 &    239$\pm$11 & 8.49$\pm$0.09 &   1.8 & A \\
40 &  082-1064 &  14:17:33.800 & 52:28:19.15 & 0.6039 &  23.26 & -18.7 &  0.41 &  1.36 &  0.18 &  -.01 &  0.06 &  40.6$\pm$0.6 &   8.2$\pm$0.8 &   5.5$\pm$0.8 &  21.1$\pm$0.8 & 8.53$\pm$0.09 &   0.4 & A \\
41 &  282-2474 &  14:15:22.654 & 52:05:04.57 & 0.6315 &  21.52 & -20.6 &  0.46 &  2.35 &  0.30 &  0.03 &  0.05 &  38.2$\pm$0.7 &  12.2$\pm$0.6 &   4.7$\pm$0.3 &  10.9$\pm$0.3 & 8.78$\pm$0.06 &   3.6 & A \\
42 &  282-3252 &  14:15:22.320 & 52:04:41.62 & 0.6317 &  22.33 & -19.8 &  0.34 &  3.09 &  0.01 &  0.08 &  0.07 &     118$\pm$1 &  36.2$\pm$0.9 &  50.1$\pm$0.5 &  141$\pm$2    & 8.47$\pm$0.05 &   4.7 & A \\
43 &  212-6648 &  14:16:03.907 & 52:12:41.35 & 0.6371 &  23.66 & -18.5 &  0.36 &  1.92 &  0.02 &  0.05 &  0.09 &     117$\pm$2 &  75.1$\pm$9.7 &     152$\pm$4 &  560$\pm$10   & 8.50$\pm$0.07 &   3.1 & A \\
44 &  093-6526 &  14:17:30.101 & 52:27:15.58 & 0.6431 &  23.45 & -18.7 &  0.38 &  1.67 &  0.37 &  0.01 &  0.04 &  63.1$\pm$0.4 &  15.3$\pm$1.5 &  15.8$\pm$0.6 &  50.3$\pm$0.6 & 8.49$\pm$0.08 &   0.7 & A \\
45 &  063-5323 &  14:17:49.728 & 52:30:31.88 & 0.6437 &  21.27 & -20.9 &  0.32 &  4.29 &  0.00 &  0.19 &  0.14 &  57.0$\pm$0.2 &  19.8$\pm$0.3 &  10.8$\pm$0.3 &  34.1$\pm$0.3 & 8.72$\pm$0.05 &   7.1 & A \\
  46 &  293-4412 &  14:15:19.779 & 52:03:30.50 & 0.6462 &  22.90 & -19.3 &  0.30 &  6.21 &  0.00 &  0.04 &  0.05 &     175$\pm$2 &  32.8$\pm$2.1 &  20.2$\pm$1.9 &  96.5$\pm$1.9 & 8.38$\pm$0.10 &   2.8 & A \\
57 &  262-3751 &  14:15:34.530 & 52:07:00.73 & 0.6497 &  20.74 & -21.3 &  0.50 &  4.68 &  0.10 &  0.18 &  0.14 &  17.4$\pm$1.5 &   5.2$\pm$0.5 &   0.5$\pm$0.4 &   4.1$\pm$0.4 & 8.85$\pm$0.07 &   3.7 & A \\
48 &  284-3046 &  14:15:30.071 & 52:03:24.75 & 0.6506 &  23.35 & -18.8 &  0.46 &  1.08 &  0.77 &  0.00 &  0.02 &  79.3$\pm$1.9 &  19.8$\pm$1.9 &  23.6$\pm$2.9 &  61.4$\pm$2.9 & 8.49$\pm$0.08 &   1.1 & A \\
49 &  172-5435 &  14:16:31.329 & 52:17:11.67 & 0.6605 &  22.55 & -19.6 &  0.48 &  3.61 &  0.01 &  0.08 &  0.03 &  49.1$\pm$0.9 &  10.3$\pm$0.9 &   4.4$\pm$2.1 &  11.8$\pm$1.9 & 8.62$\pm$0.10 &   1.3 & A \\
50 &  152-1633 &  14:16:48.315 & 52:19:38.62 & 0.6736 &  21.80 & -20.5 &  0.44 &  6.99 &  0.02 &  0.06 &  0.09 & 119$\pm$2     &  27.0$\pm$0.9 &  14.3$\pm$5.7 &    44$\pm$5.7 & 8.56$\pm$0.09 &   7.2 & A \\
51 &  063-7209 &  14:17:47.868 & 52:30:47.38 & 0.6760 &  21.01 & -21.3 &  0.33 &  3.73 &  0.49 &  0.09 &  0.17 &  69.3$\pm$0.3 &  34.1$\pm$0.3 &  16.1$\pm$0.9 &  50.8$\pm$1.7 & 8.80$\pm$0.04 &  17.2 & A \\
52 &  164-3859 &  14:16:48.501 & 52:17:15.55 & 0.6828 &  22.42 & -19.9 &  0.47 &  4.62 &  0.02 &  0.03 &  0.05 &  64.1$\pm$1.9 &  14.5$\pm$1.1 &   3.9$\pm$1.9 &  11.5$\pm$1.9 & 8.65$\pm$0.09 &   2.3 & A \\
53 &  292-4369 &  14:15:14.367 & 52:03:45.35 & 0.6926 &  20.61 & -21.7 &  0.47 &  3.41 &  0.58 &  0.07 &  0.09 &  39.3$\pm$0.5 &  22.2$\pm$1.7 &  11.5$\pm$3.7 &  34.6$\pm$3.7 & 8.83$\pm$0.04 &  19.6 & A \\
54 &  282-6050 &  14:15:19.468 & 52:04:32.81 & 0.6946 &  22.70 & -19.7 &  0.39 &  2.50 &  0.01 &  0.08 &  0.12 &  97.9$\pm$1.8 &  35.4$\pm$1.8 &  29.4$\pm$2.8 &   103$\pm$3.0 & 8.61$\pm$0.05 &   4.3 & A \\
55 &  153-3721 &  14:16:51.198 & 52:19:47.44 & 0.7043 &  21.07 & -21.4 &  0.32 &  3.19 &  0.25 &  0.10 &  0.08 &  71.5$\pm$0.9 &  32.0$\pm$0.9 &  23.4$\pm$0.9 &    71$\pm$0.9 & 8.72$\pm$0.04 &  16.9 & A \\
56 &  292-7343 &  14:15:11.836 & 52:03:14.00 & 0.7450 &  23.56 & -19.0 &  0.29 &  1.64 &  0.02 &  0.03 &  0.07 &  99.1$\pm$1.8 &  24.1$\pm$4.4 &  26.3$\pm$4.5 &  79.6$\pm$3.6 & 8.46$\pm$0.12 &   1.4 & A \\
57 &  084-6809 &  14:17:42.554 & 52:27:29.20 & 0.7467 &  21.92 & -20.6 &  0.46 &  6.54 &  0.09 &  0.10 &  0.12 &     121$\pm$2 &  26.2$\pm$1.8 &  16.0$\pm$2.0 &    48$\pm$2.0 & 8.52$\pm$0.09 &   8.0 & A \\
58 &  182-7536 &  14:16:22.681 & 52:15:59.33 & 0.7505 &  22.73 & -19.8 &  0.40 &  3.86 &  0.04 &  0.07 &  0.08 &  72.5$\pm$1.8 &  18.2$\pm$1.2 &  13.7$\pm$1.4 &  37.9$\pm$1.4 & 8.57$\pm$0.07 &   2.4 & A \\
59 &  292-6724 &  14:15:12.920 & 52:02:56.62 & 0.7659 &  21.93 & -20.7 &  0.47 &  3.37 &  0.09 &  0.06 &  0.10 &  35.3$\pm$0.9 &  11.3$\pm$0.9 &   4.2$\pm$1.6 &  11.8$\pm$1.6 & 8.77$\pm$0.07 &   3.7 & A \\
60 &  134-0967 &  14:17:05.035 & 52:20:31.42 & 0.7885 &  22.83 & -19.9 &  0.26 &  3.76 &  0.00 &  0.14 &  0.07 &     117$\pm$2 &  51.9$\pm$1.9 &  58.1$\pm$2.5 &  225$\pm$3    & 8.55$\pm$0.04 &   6.9 & A \\
61 &  142-4838 &  14:16:51.266 & 52:20:45.96 & 0.8077 &  20.18 & -22.6 &  0.28 &  5.17 &  0.25 &  0.16 &  0.09 &  13.1$\pm$1.9 &  11.4$\pm$0.5 &   9.5$\pm$0.7 &  33.3$\pm$0.5 & 8.80$\pm$0.02 &  18.5 & A \\
62 &  174-3527 &  14:16:40.994 & 52:16:35.72 & 0.8096 &  21.59 & -21.2 &  0.46 &  3.19 &  0.00 &  0.10 &  0.18 &  56.9$\pm$1.7 &  25.8$\pm$3.1 &  10.4$\pm$2.1 &  23.9$\pm$2.1 & 8.84$\pm$0.06 &  13.2 & A \\
63 &  152-3226 &  14:16:46.841 & 52:19:28.70 & 0.8127 &  22.52 & -20.3 &  0.34 &  4.42 &  0.42 &  0.05 &  0.10 &     109$\pm$2 &  28.3$\pm$4.4 &  22.9$\pm$3.0 &  88.8$\pm$3.0 & 8.50$\pm$0.09 &   5.4 & A \\
64 &  153-6078 &  14:16:56.673 & 52:20:22.44 & 0.8128 &  20.98 & -21.8 &  0.48 &  6.87 &  0.06 &  0.14 &  0.15 &  70.0$\pm$5.2 &  59.5$\pm$7.9 &   7.7$\pm$1.5 &  23.4$\pm$1.5 & 9.00$\pm$0.03 &  58.3 & A \\
65 &  092-7832 &  14:17:20.880 & 52:26:23.36 & 0.6824 &  23.57 & -18.7 &  0.29 &  1.43 &  0.54 &  0.00 &  0.10 &  59.0$\pm$1.9 &   154$\pm$0.7 &   117$\pm$4   &   237$\pm$4   & 8.93$\pm$0.04 &   8.2 & D \\
66 &  203-3109 &  14:16:17.618 & 52:13:49.43 & 0.6848 &  22.41 & -19.9 &  0.34 &  5.60 &  0.02 &  0.10 &  0.08 &  60.5$\pm$1.7 &  30.2$\pm$1.5 &  11.3$\pm$2.2 &  31.1$\pm$2.2 & 8.85$\pm$0.04 &   4.9 & D \\

\\
\\
\\
\\
\\
\\
\tablebreak

\enddata
\tablerefs{
(1) Reference ID \# for this paper;
(2) Groth Strip survey ID ; 
(3) J2000 Right Ascension; 
(4) J2000 Declination; 
(5) redshift from optical emission lines;
(6) $I_{814}$ AB magnitude;
(7) rest-frame $M_B$ for $H_0=70$ km/s/Mpc, $\Omega_M=0.3$, $\Omega_\Lambda=0.7$;
(8) rest-frame B-V color;
(9) half light radius in kpc derived from the global $I_{814}$ image;
(10) bulge fraction derived from the $I_{814}$ image;
(11) $R_T$ asymmetry index as defined in paper II;
(12) $R_A$ asymmetry index as defined in Paper II;
(13) measured EW of [O~II] $\lambda$3727 and uncertainty, corrected to the rest frame;
(14) measured EW of H$\beta$ $\lambda$4861 and uncertainty, corrected to the rest frame 
	(no correction for stellar absorption);
(15) measured EW of [O~III] $\lambda$4959 and uncertainty, corrected to the rest frame;
(16) measured EW of [O~II] $\lambda$5007 and uncertainty, corrected to the rest frame;
(17) oxygen abundance, 12+log(O/H), and 1$\sigma$ uncertainty from the empirical $R_{23}$ method
following McGaugh (1991) as formulated in KKP after correction for 2 \AA\ of 
stellar absorption in the $EW_{H\beta}$.  An additional uncertainty of 20\% has been
added to the $EW_{[O~II]}$ as discussed in the text.  An additional uncertainty of 
$\sim$0.15 dex in O/H representing uncertainties in the
photoionization models and empirical strong-line
calibration should be added in quadrature to the tabulated
measurement errors;
(18) estimated star formation rate based on $H\beta$ flux 
(not corrected for stellar absorption), derived from the
	EW($H\beta$) and V-band absolute magnitude.  We estimate the
	$H\beta$ luminosity as $L_{H\beta}(erg/s)=5.49\times10^{31} \times 2.5^{-M_V} \times EW_{H\beta}$.
	The SFR then is computed by
	$SFR(M_\odot/yr)= 2.8 \times L_{H\beta} /1.12\times10^{41}$
	which assumes the Kennicutt (1983) calibration of SFR in terms
	of $H\alpha$ luminosity.  This estimate is a lower limit since extinction
	and stellar Balmer absorption is not taken into account;  
(19) Notes: 
A--Probable star-forming galaxy (non-AGN or minimal AGN contribution), as defined in text;
B--[O~II] EW is particularly uncertain due to low continuum levels; 
C--Low-luminosity object with line ratios in the
``turn-around'' region of the strong-line abundance
diagram, so the oxygen abundance is 
highly uncertain, in the range $7.8<12+log(O/H)<8.4$.  Object removed
from remaining analysis.
D--Object is a probable AGN on basis of [Ne III]/[O~II] $>0.4$.}
\end{deluxetable}

\begin{deluxetable}{lccrrrr}
\label{fittable.tab}
\tabletypesize{\scriptsize} 
\setlength{\tabcolsep}{0.02in} 
\tablewidth{3.9in}
\tablecaption{L-Z Relation Fits: $12+log(O/H)=X\times M_B + Y $}
\tablehead{
\colhead{Sample} & 
\colhead{z} & 
\colhead{\#Galaxies} & 
\colhead{X}   & 
\colhead{Y}   &
\colhead{$\Delta M_B$}   &
\colhead{$\Delta O/H$ }  \\ [.2ex]
\colhead{(1)} & 
\colhead{(2)} & 
\colhead{(3)} & 
\colhead{(4)} & 
\colhead{(5)} &
\colhead{(6)} &
\colhead{(7)} }
\startdata
K92+NFGS+KISS & $<$0.09   & 133  & -0.063 (0.007) & 7.60 & +1.0 & +0.07 \\
K92+NFGS      & $<$0.09   & 54   & -0.052 (0.009) & 7.74 &  0.0 & 0.0  \\
DGSS          & 0.29-0.40 & 15   & -0.056 (0.024) & 7.56 & -1.6 & -0.10 \\
DGSS+KZ99     & 0.40-0.60 & 24   & -0.082 (0.014) & 7.02 & -2.0 & -0.12 \\
DGSS+CL01     & 0.60-0.82 & 33   & -0.128 (0.012) & 6.04 & -2.4 & -0.15 \\
\enddata
\tablerefs{
(1) galaxy sample, by redshift;
(2) redshift range;
(3) number of galaxies in sample;
(4) slope of metallicity-luminosity relation and uncertainty;
(5) intercept of the unweighted least squares fit;
(6) offset, in magnitudes, of the fit compared to the local K92+NFGS sample at the approximate midpoint of the 
	$M_B$ distributions, $M_B=-20.5$;
(7) offset, in dex, of the fit compared to the local K92+NFGS sample at the approximate midpoint of the 
	O/H distributions, 12+log(O/H)=8.7.}
\end{deluxetable}

\begin{deluxetable}{lrr}
\label{costable.tab}
\tabletypesize{\scriptsize} 
\setlength{\tabcolsep}{0.02in} 
\tablewidth{1.8in}
\tablecaption{Lookback Time and Redshift}
\tablehead{
\colhead{z} & 
\colhead{$T_{lookback}$} & 
\colhead{$Age_{universe}$}   \\ [.2ex]
\colhead{(1)} & 
\colhead{(2)} & 
\colhead{(3)}  }
\startdata
0.1 & 1.30 & 12.15 \\
0.2 & 2.43 & 11.02 \\
0.4 & 4.28 & 9.17 \\
0.6 & 5.71 & 7.74 \\
0.8 & 6.82 & 6.62 \\
1.6 & 9.49 & 3.95 \\
3.2 & 11.50 & 1.95 \\
6.4 & 12.61 & 0.83 \\
10 & 13.0 & 0.45 \\
\enddata
\tablerefs{
(1) Redshift;
(2) Lookback time in Gyr for $H_0$=70 \kms\ Mpc$^{-1}$, $\Omega_m=0.3$, and $\Omega_\Lambda=0.7$
(3) Age of universe in Gyr
}
\end{deluxetable}

\end{document}